\documentclass[floatfix,nofootinbib,twocolumn,preprintnumbers,superscriptaddress,notitlepage]{revtex4-1}

\usepackage{times}
\usepackage{xcolor}
\usepackage{amsfonts}
\usepackage[tbtags]{amsmath}
\usepackage{amssymb}
\usepackage{bm}
\usepackage{dcolumn}
\usepackage{enumerate}
\usepackage{epsfig}
\usepackage{graphicx}
\usepackage{graphics}
\usepackage[latin1]{inputenc}
\usepackage{latexsym}
\usepackage{rotating}
\usepackage{hyperref}
\usepackage[caption=false]{subfig}
\usepackage{longtable}
\usepackage{supertabular}
\usepackage[normalem]{ulem}

\newcommand{\<}{\begin{equation}}
\newcommand{\?}{\end{equation}}

\definecolor{mgreen}{rgb}{0.1,0.7,0.1}

\begin{document}

\title{Numerical Relativity Estimates of the Remnant Recoil Velocity in Binary Neutron Star Mergers}

\author{Sumeet Kulkarni}
\affiliation{Department of Physics and Astronomy, The University of Mississippi, University, Mississippi 38677, USA}
\author{Surendra Padamata}
\affiliation{Department of Physics, The Pennsylvania State University, University Park, PA 16802, USA}
\author{Anuradha Gupta}
\affiliation{Department of Physics and Astronomy, The University of Mississippi, University, Mississippi 38677, USA}
\author{David Radice}
\affiliation{Department of Physics, The Pennsylvania State University, University Park, PA 16802, USA}
\affiliation{Institute for Gravitation and the Cosmos, The Pennsylvania State University, University Park, PA 16802, USA}
\affiliation{Department of Astronomy \& Astrophysics, The Pennsylvania State University, University Park, PA 16802, USA}
\author{Rahul Kashyap}
\affiliation{Department of Physics, The Pennsylvania State University, University Park, PA 16802, USA}
\affiliation{Institute for Gravitation and the Cosmos, The Pennsylvania State University, University Park, PA 16802, USA}

\date{\today}

\begin{abstract}
We present, for the first time, recoil velocity estimates for binary neutron star mergers using data from numerical relativity simulations. 
We find that binary neutron star merger remnants can have recoil velocity of the order of a few tens of km/s and as high as $150$ km/s in our dataset. These recoils are attained due to equivalent contributions from the anisotropic gravitational wave emission as well as the asymmetric ejection of dynamical matter during the merger. We provide fits for net recoil velocity as well as its ejecta component as a function of the amount of ejected matter, which may be useful when constraints on the ejected matter are obtained through electromagnetic observations. 
We also estimate the mass and spin of the remnants and find them to be in the range $[2.34, 3.38] M_{\odot}$ and $[0.63, 0.82]$ respectively, for our dataset.
 
\end{abstract}

\maketitle

\section{Introduction}
\label{sec:intro}
The LIGO and Virgo~\cite{LIGOScientific:2014pky, VIRGO:2014yos} ground-based gravitational wave (GW) detectors have so far confidently detected signals from two binary neutron stars (BNSs): GW170817~\cite{LIGOScientific:2017vwq} and GW190425~\cite{LIGOScientific:2020aai}.
The first event, GW170817, was the merger of two neutron stars with a total mass of~${\sim} 2.73 M_{\odot}$. Its resulting electromagnetic counterpart and kilonova were also observed by several telescopes~\cite{LIGOScientific:2017ync}, enabling us to study this event with multiple messengers. Despite this, the very nature of the GW170817 merger remnant is still uncertain~\cite{LIGOScientific:2018urg, LIGOScientific:2019eut, Soma:2019utv}. The second event, GW190425, had a higher total mass of ${\sim} 3.4 M_{\odot}$, which in addition to the non-detection of its electromagnetic counterpart, means that this event retains the plausibility of being a neutron star-black hole merger~\cite{Han:2020qmn, Kyutoku:2020xka}. While one can predict the energy and angular momentum at the end of BNS inspiral fairly accurately~\cite{Bernuzzi:2020tgt}, the final mass and spin of the remnants have large  uncertainties because BNSs radiate a considerable amount of energy post-merger, especially if they form a hypermassive neutron star~\cite{Bernuzzi:2015opx}. The dynamical ejection of matter towards the end of the merger, and the resulting interaction of the remnant with its outer ejecta disk can also influence its properties. The computational demands of studying BNS mergers using numerical relativity (NR) and relativistic hydrodynamic simulations have led to insufficient constraints on their remnant properties for a broad range of systems. In particular, we lack knowledge about the recoil velocities (or ``kicks") of BNS merger remnants altogether.

It is important to estimate the recoils imparted in BNS merger remnants because of their implications in hierarchical mergers in different astrophysical environments. 
The dynamical formation scenario~\cite{PortegiesZwart:1999nm, Rodriguez:2015oxa} involves compact objects in dense environments to form binaries via dynamical interactions. These may be first-generation compact objects born out of independent core-collapse supernovae to form first-generation mergers. But this scenario also includes the feature of hierarchical mergers, where remnants of first-generation mergers 
are retained in the environment and are available for successive mergers. The crucial factor for making higher-generation binary mergers possible is the magnitude of the remnant's recoil in relation to the escape speed of the environment. 
Recoils higher than the environment's escape speed deem the remnant unavailable for hierarchical mergers. In turn, if dynamical mergers are more prevalent, the distribution of recoil may determine the merger rate of binaries and the nature of the retained compact object that participate in these mergers~\cite{Doctor:2019ruh, Gupta:2019nwj, Gerosa:2019zmo, Kimball:2020opk, Mahapatra:2021hme}.

Remnant properties of BBH mergers are very well studied, thanks to many NR simulations performed over the years for binary black holes (BBHs)~\cite{Mroue:2013xna,Jani:2016wkt,Boyle:2019kee,Healy:2022wdn}. 
Consequently, we have reliable and accurate predictions
for properties of the BBH merger remnants such as their final mass, spin and
recoil velocity through analytical~\cite{Damour:2006tr, Sopuerta:2006wj}, semi-analytical~\cite{Deng:2020rnf} models as well as NR simulations~\cite{Campanelli:2007cga, Lousto:2009mf, Healy:2014yta, CalderonBustillo:2018zuq, Varma:2018aht, Varma:2020nbm}.
While there are some predictions for the final mass and spin of
BNS merger remnants~\cite{Piro:2017zec, Zappa:2017xba,Deng:2020rnf},
no NR-based estimates exist in the literature for the remnant's recoil velocity. Therefore, in this paper, we present estimates of the recoil velocities of BNS merger remnants computed using numerical simulations published in~\cite{Gonzalez:2022mgo} that were produced using the WhiskyTHC code \cite{Radice:2012cu,Radice:2013hxh,Radice:2013xpa}, and are part of the Computational Relativity (\textsc{CoRe}) database.
Typically, the primary way of calculating BBH recoils involves an estimation of the net linear momentum emitted in a preferred direction by GWs. However, in BNS mergers, asymmetric dynamical ejection of matter at high velocities can also impart a recoil on the central remnant. In this work, we calculate both the GW and ejecta emission components of the recoils (see Sec.~\ref{sec:methods}). 
We discuss the dependence of these recoils on various binary parameters, the neutron star equation of state (EoS), as well as the properties of dynamically ejected matter. We also provide estimates for BNS remnant's mass and spin.

Our main finding is that BNS remnants can get significantly large kicks due to dynamical ejection of matter. This leads to the net remnant recoils exceeding $100$ km/s in some cases. We compare the contribution of GW and ejecta emission towards the net remnant recoils and find them to be comparable for a majority of cases. We also report a dependence of these recoils on the amount and asymmetry of ejected matter and calculate an empirical fitting formula for recoil velocity as a function of the net amount of ejecta matter.

This paper is organized as follows: Sec.~\ref{sec:data} briefly describes the parameter space of our NR dataset and gives details of the general relativistic-hydrodynamic (GRHD) simulations. Section~\ref{sec:methods} explains how we estimate the recoil for each merger remnant due to the two mechanisms, viz. the GW emission and dynamical ejecta. The distribution of recoils for our BNS simulations is presented and analyzed in Sec.~\ref{sec:results}, where we also compare the two recoil mechanisms and discuss the physical insights the data offers in explaining their role. In Sec.~\ref{sec:discussion} we discuss various limitations of this work and provide the astrophysical implications of our results.

\section{GRHD simulations}
\label{sec:data}

\begin{figure*}[!ht]
\centering
\subfloat{
\includegraphics[width=0.33\textwidth,trim = {0 0 0 10}]{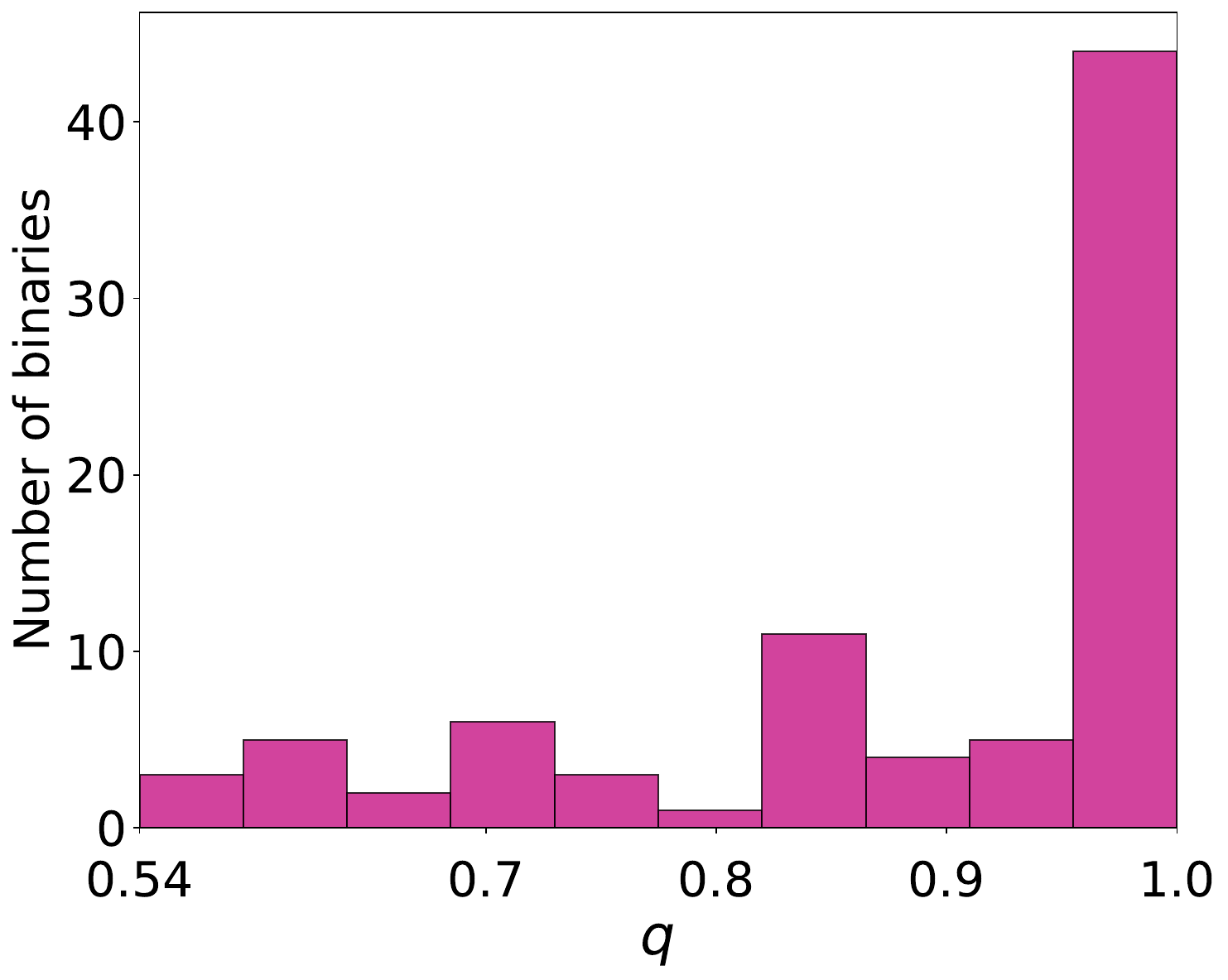}
}
\subfloat{
\includegraphics[width=0.33\textwidth,trim = {0 0 0 10}]{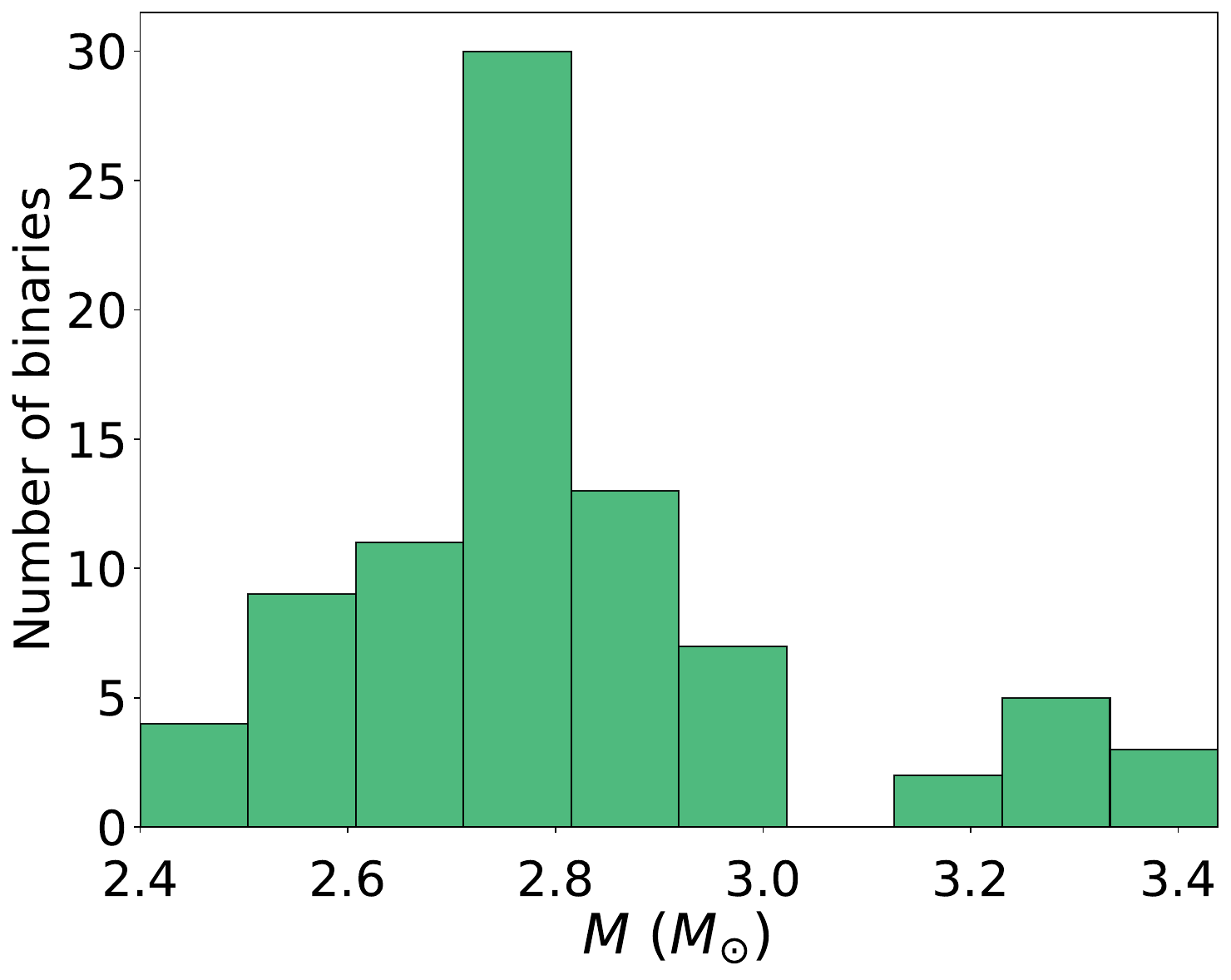}
}
\subfloat{
\includegraphics[width=0.33\textwidth,trim = {0 0 0 10}]{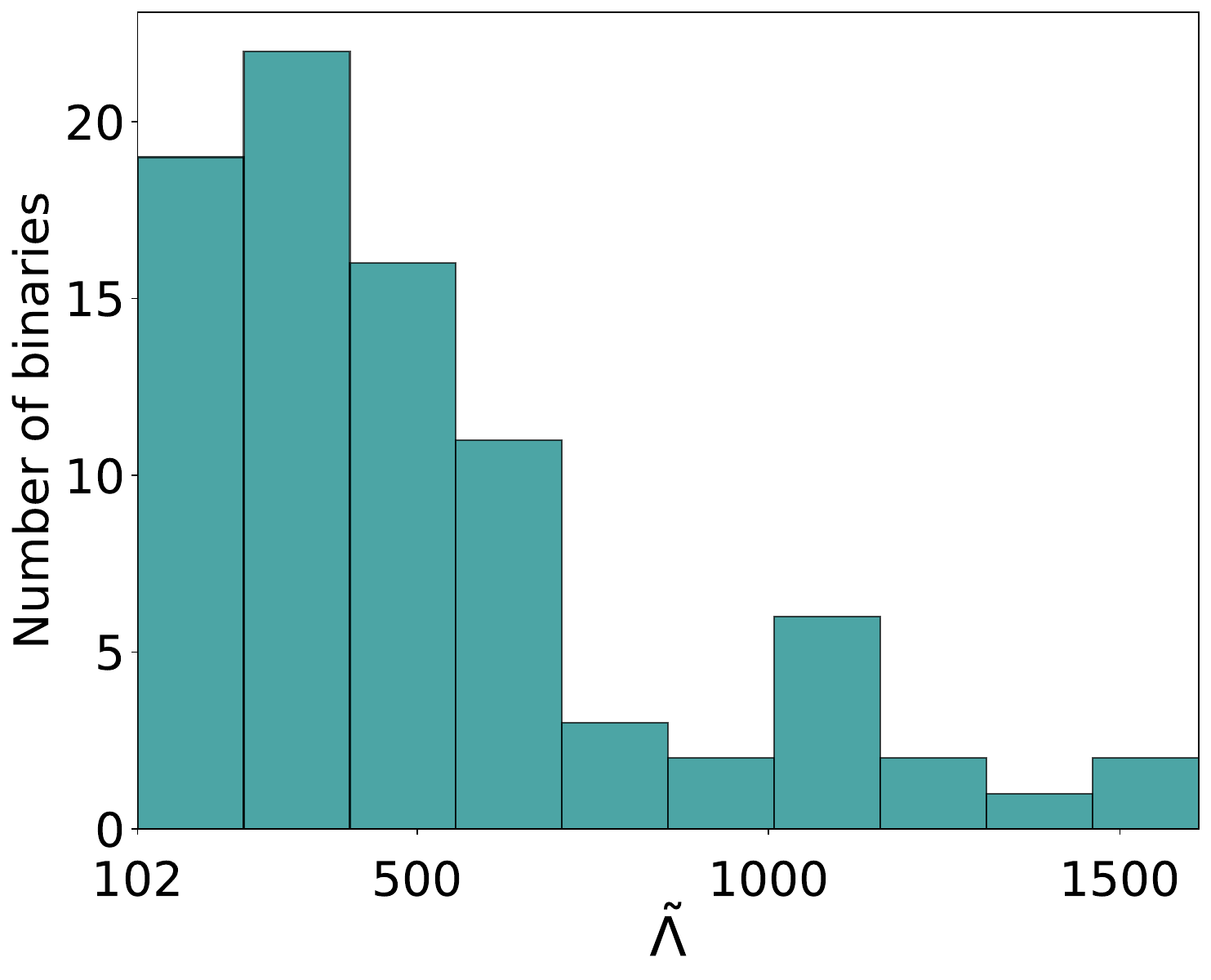}
}
\caption{\label{fig:dataset-params} The distribution of mass ratio ($q$), total mass ($M$) and reduced tidal deformability parameter ($\Tilde{\Lambda}$) for the 84 binaries in our dataset. 
}
\end{figure*}

The dataset we considered includes a total of $200$ NR simulations of BNS mergers taken from the \textsc{CoRe} database \cite{Gonzalez:2022mgo}. We only used the simulations that were run using the  WhiskyTHC code
\cite{Radice:2012cu,Radice:2013hxh,Radice:2013xpa}. 
Figure~\ref{fig:dataset-params} shows the distribution of mass ratio ($q=m_2/m_1 \in (0,1]$), total mass ($M=m_1+m_2$), and the reduced tidal deformability parameter ($\Tilde{\Lambda}$) of the binaries in our dataset. Here $m_1$ and $m_2$ are the neutron star masses in the binary ($m_1$ being the primary) and $\Tilde{\Lambda}$ is defined as~\cite{Wade:2014vqa}:
\begin{equation*}
    \widetilde{\Lambda} = \frac{16}{13} \frac{(m_{1}+12m_{2})m_{1}^{4}\Lambda_{1} + (m_{2}+12m_{1})m_{2}^{4}\Lambda_{2}}{(m_{1}+m_{2})^{5}},
\end{equation*}
with $\Lambda_{1}$ and $\Lambda_{2}$ being the tidal deformability of two neutron stars. All binary neutron star systems are irrotational which implies component neutron stars are essentially non-spinning in our dataset. This is a reasonable approximation as typical galactic BNS systems merging within Hubble time are expected to have low spins prior to the merger. This approximation can be obtained by extrapolating the neutron star's spin-down rate till merger \cite{Burgay:2003jj, LIGOScientific:2017vwq}. 
Explicitly, the ranges of $q$, $M$ and $\tilde \Lambda$ in our dataset are as follows: $q = [0.55,1]$, $M = [2.40, 3.44] M_{\odot}$, $\tilde \Lambda = [101.72, 1612.24]$. 

The initial data for these simulations are constructed as quasi-circular binary using the multi-domain spectral code Lorene \cite{lorene_url} based on the extended thin sandwich formalism \cite{Gourgoulhon:2007ue}. The initial separation of the binary is typically 40 km. Z4c~\cite{Bruegmann:2003aw,Bernuzzi:2009ex,Hilditch:2012fp} hyperbolic conformal formulations of 3+1 Einstein field equations are used for evolving the initial data along with a Kurganov-Tadmor type scheme with high-order primitive reconstruction with MP5 and second-order accurate flux integration. Refluxing was used in all simulations. The setup of these simulations is identical to~\cite{Radice:2018pdn}. Carpet adaptive mesh refinement framework that implements Berger-Oliger scheme with 7 level mesh refinement was implemented. Different fine grid resolutions, in particular, a low resolution (in geometric units) of `0.167$M_{\odot}$'$\approx 246$ m, a standard resolution of `0.125$M_{\odot}$'$\approx 185$ m, and a high resolution of `0.083$M_{\odot}$'$\approx 123$ m were used. Note here that not all binaries have been simulated at all the above resolutions. Gravitational wave extraction is performed by computing the Weyl psuedoscalar $\Psi_{4}$ at an extraction radius, $r = 400M_{\odot}$, following the method employed in the open source software Einstein Toolkit~\cite{EinsteinToolkit:web}. Most simulations used the M0~\cite{Radice:2016dwd} and, a few, leakage scheme \cite{Radice:2016dwd,Galeazzi:2013mia} for the neutrino transport.

\begin{figure}[h]
\centering
\subfloat{
\includegraphics[width=0.95\linewidth]{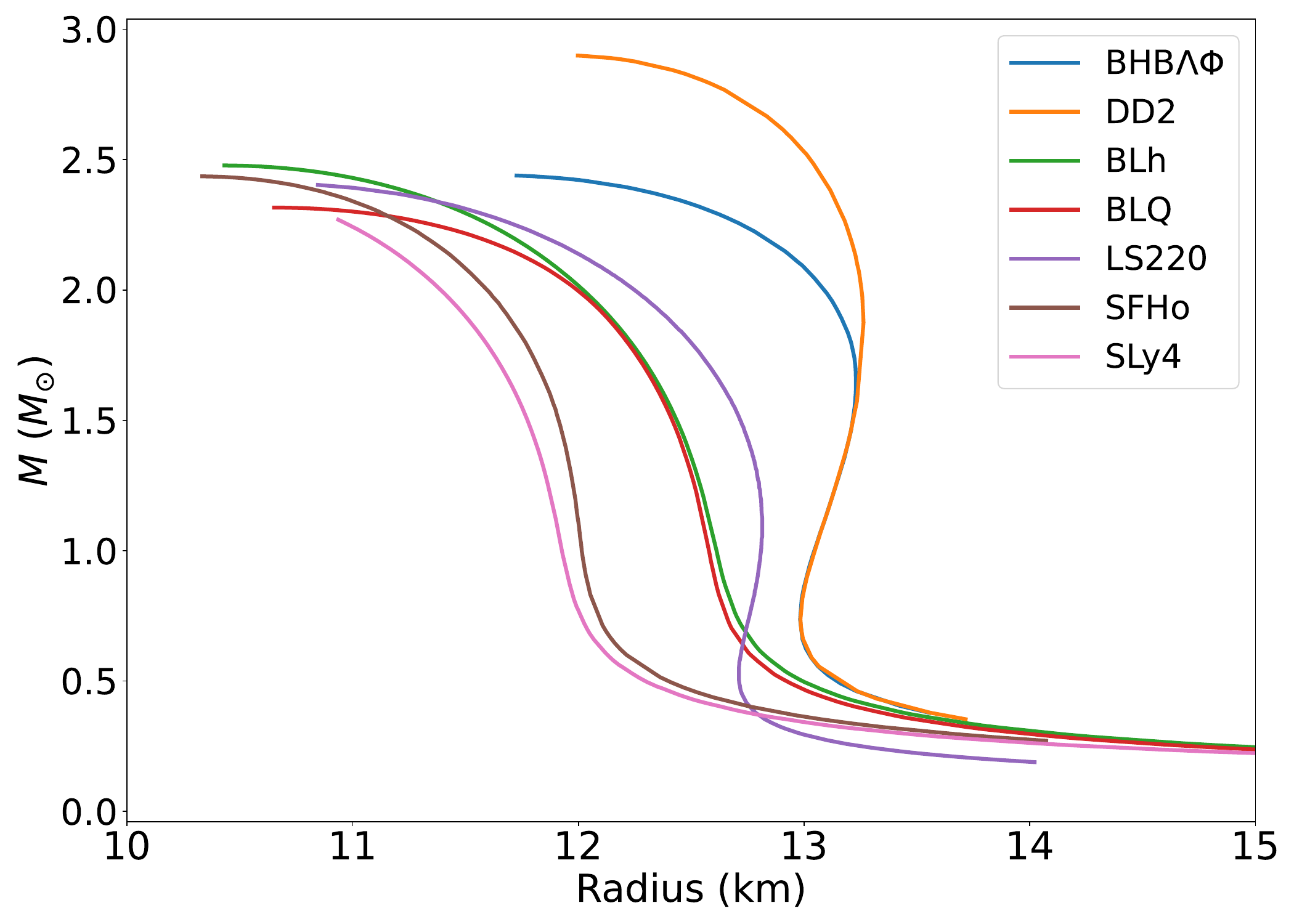}}
\caption{\label{fig:NS_EOS} The neutron star equations of state used in our numerical relativity simulations.  
}
\end{figure}

Various microphysical EoSs which are finite-temperature and have the electrons, neutrons, protons, nuclei, positrons, and photons distributions as their degrees of freedom were used. The EoSs are BHB$\Lambda\phi$~\cite{Banik:2014qja}, BLh~\cite{Logoteta:2020yxf,Bombaci:2018ksa}, BLQ~\cite{Prakash:2021wpz,Logoteta:2020yxf}, DD2~\cite{Typel:2009sy,Hempel:2009mc}, LS220~\cite{1991NuPhA.535..331L}, SFHo~\cite{Steiner:2012rk}, and SLy4~\cite{Douchin:2001sv,Schneider:2017tfi}. BLQ is a hybrid model based off of BLh, where states with deconfined quarks are possible. BHB$\Lambda\phi$ is a hadronic EoS that has an additional degree of freedom, $\Lambda$ hadrons \cite{Banik:2014qja,Radice:2016rys}. The mass-radius curves corresponding to these EoSs are shown in Fig.~\ref{fig:NS_EOS}. For further description of the simulations used here, see~\cite{Gonzalez:2022mgo}. 

During the binary evolution, close to the merger, matter is ejected due to tidal interactions and shocks between the two neutron stars and some of it becomes unbound, which is dubbed as \emph{dynamical outflow} in literature \cite{PhysRevD.87.024001, 2013ApJ...773...78B, Radice:2016dwd, Radice:2018pdn}. More ejecta could become unbound due to neutrino winds and on longer timescales in post-merger due to magnetic effects and nuclear recombination as well, usually called \emph{secular ejecta}. But we only consider the above-mentioned dynamical outflow in estimating the recoil of the remnant and do not include the \emph{secular ejecta} or any other winds' contribution. We estimate the rest mass of the outflow using the geodesic criteria, integrating in time flux of matter with time component of four velocity, $u_{t} < -1$ over a 2-sphere with radius $200 M_{\odot} \approx 295$ km. For more details and various studies on the dynamical outflow, we refer the reader to \cite{Radice:2018pdn,Henkel:2022naw}. Note that the ejecta data is not publicly available in the \textsc{CoRe} database but can be made available upon request. 

Since the total dataset contains degenerate runs corresponding to any given binary due to multiple resolutions and neutrino transport schemes, we select cases with the highest possible resolution and the M0 neutrino transport mechanism whenever available for each unique binary defined by $M$, $q$, and the EoS. This reduces the size of our dataset from 200 to $84$ binaries.

\section{Method to Compute BNS merger Remnant Properties}
\label{sec:methods}
In this section, we provide the methods of how we estimate BNS merger remnant properties using NR simulations. 
For each BNS simulation in our database, we calculate separately the recoil due to GW emission (Sec.~\ref{ssec:gw_recoil}) and ejecta emission (Sec.~\ref{ssec:ejecta_recoil}). 
We also calculate the mass and spin of the remnants as described in Sec.~\ref{ssec:mass_spin}.

\subsection{Calculation of Gravitational-Wave Recoils}
\label{ssec:gw_recoil}
Mass and spin asymmetries in the binary cause the emitted GWs to carry away linear momentum in some preferred direction~\cite{Favata:2004wz}. 
As a result of the conservation of linear momentum, a recoil is imparted to the binary's center of mass in the opposite direction. These recoils are negligible for non-spinning systems with equal component masses~\cite{Koppitz:2007ev}. 
The components of astrophysical BNS systems are expected to have very low spins~\cite{LIGOScientific:2017vwq} and comparable masses ($q \sim 1$) owing to the limited range of neutron star masses. Therefore, BNS recoils are expected to be small due to GW emission.

In order to compute the recoils for binaries in the \textsc{CoRe} NR database, we follow the treatment described in~\cite{Gerosa:2018qay, Ruiz:2007yx} where the net emission of linear momentum can be derived from the NR-based GW strain $h(t)$ (at the extraction radius $r$) expressed in terms of spherical harmonics as follows: 

\begin{equation}
\label{eq:hoft}
	\begin{split}
h(t, \theta, \phi, \pmb{\lambda}) &= 
h_{+}(t, \theta, \phi, \pmb{\lambda}) - ih_{\times}(t, \theta, \phi, \pmb{\lambda}) \\ 
&= \sum_{l=2}^{\infty}\sum_{m=-l}^{+l} h^{lm}(t,\pmb{\lambda})_{-2} Y^{lm}(\theta, \phi), 
	\end{split}
\end{equation}
where $(\theta, \phi)$ are the right ascension and declination sky location coordinates, while $\pmb{\lambda}$ represents all intrinsic parameters of the binary. The emission of linear momentum along the three cartesian components ($P_{x}, P_{y}, P_{z}$) can be obtained by summing over the different $(l,m)$ modes of $\dot{h}(t)$ at the extraction radius $r$:

\begin{subequations}
\label{eq:dpdt2}
\begin{align}
\begin{split}
\frac{dP_{+}}{dt} &= \frac{1}{8\pi} \sum_{l,m}\dot{h}^{l,m} \Big[a_{l,m}\dot{h}^{l,m+1} + b_{l,-m}\dot{h}^{l-1,m+1} \\ 
&-  b_{l+1,m+1}\dot{h}^{l+1,m+1} \Big],
\end{split}\\
\begin{split}
\frac{dP_{z}}{dt} &= \frac{1}{16\pi} \sum_{l,m}\dot{h}^{l,m} \Big[c_{l,m}\dot{h}^{l,m} + d_{l,m}\dot{h}^{l-1,m} \\ 
&- b_{l+1,m}\dot{h}^{l+1,m} \Big],
\end{split}
\end{align}
\end{subequations}
where $P_{+} = P_{x}+iP_{y}$, and the spectral coefficients are given by:
\begin{subequations}
\begin{align}
\begin{split}
a_{l,m} &= \frac{\sqrt{(l-m)(l+m+1)}}{l(l+1)}, 
\end{split} \\
\begin{split}
b_{l,m} &= \frac{1}{2l} \sqrt{\frac{(l-2)(l+2)(l+m)(l+m+1)}{(2l-1)(2l+1)}}, 
\end{split} \\
\begin{split}
c_{l,m} &= \frac{2m}{l(l+1)},
\end{split} \\
\begin{split}
d_{l,m} &= \frac{1}{l} \sqrt{\frac{(l-2)(l+2)(l+m)(l+m)}{(2l-1)(2l+1)}}.
\end{split}
\end{align}
\end{subequations}

The total radiated momentum can be obtained by integrating Eqs.~(\ref{eq:dpdt2}) over time to obtain the GW recoil velocity:

\begin{equation}
\label{eqn:voft}
\mathbf{v_{\rm rec, gw}} = - \frac{P_{x}\mathbf{\hat{x}}+P_{y}\mathbf{\hat{y}}+P_{z}\mathbf{\hat{z}}}{M_{\rm rem}}.
\end{equation}
Here $P_x$, $P_y$, $P_z$ are the final values of the Cartesian components of radiated linear momenta at the end of the simulation, and $M_{\rm rem}$ is the total mass of the system after subtracting 
the energy radiated due to GW emission as well as the mass loss due to ejecta, as defined below in Eq.~(\ref{eq:M_rem}). The mass loss due to GW emission can be computed by integrating the GW luminosity expression: 

\begin{equation}
\label{eq:Erad}
\dot{E}_{\rm GW} = \frac{dE_{\rm GW}}{dt} = \lim_{t\to\infty} \frac{r^{2}}{16\pi} |\dot{h}^{l,m}|^{2}.
\end{equation}

\subsection{Calculation of Recoils due to Ejecta Emission}
\label{ssec:ejecta_recoil}
In our dataset, a typical merger of neutron stars results in a dynamical ejection of matter as discussed in Sec.~\ref{sec:data}, with an average of $0.06 \%$ of the total mass of the system being ejected. This ejecta is thrown out at high speeds, with the root-mean-squared values of the ejecta velocity distributions for the binaries lying in the range [$0.19c, 0.75c$], with $c$ being the speed of light.
If the distribution of this ejecta is asymmetric about the center of mass, the resulting compact object will experience a recoil due to the conservation of linear momentum. We employ a Newtonian picture while dealing with ejecta as the bulk of it is not relativistic and the metric corrections would be small. Further, we assume that the baryonic and gravitational mass losses due to ejecta are the same.

We calculate the net momentum of the ejecta by integrating the time-averaged ejecta momentum flux density distribution over a 2-sphere, analogous to the determination of neutron star supernova natal kick as discussed in~\cite{Janka:2016nak}:

\begin{equation}
\label{eq:P_ej}
\mathbf{P}_{\rm ej} = \int dt \int_{S}~\rho~\mathbf{v}~(\mathbf{v} \cdot d\mathbf{S}),
\end{equation}
where integration is carried out over the extraction sphere $S$. Here $\rho$ and $\mathbf{v}$ are the ejecta mass density and ejecta velocity respectively. The net mass carried out by the outflow can be similarly determined as: 
\begin{equation}
\label{eq:M_ej}
M_{\rm ej} = \int dt \int_{S} \rho~(\mathbf{v} \cdot d\mathbf{S}).
\end{equation}
Subtracting this ejected mass ($M_{\rm ej}$) along with mass loss due to emission of GW radiation gives us the final remnant mass, $M_{\rm rem}$ [defined below in Eq.~(\ref{eq:M_rem})]. Conservation of linear momentum then implies,
\begin{equation}
\mathbf{v}_{\rm rec, ej} = -\frac{\mathbf{P}_{\rm ej}}{M_{\rm rem}},
\end{equation}
with $\mathbf{v}_{\rm rec, ej}$ being the recoil imparted due to ejecta emission.

We take the net recoil imparted to the BNS merger remnant as the vector sum of the GW and ejecta recoils:

\begin{equation}
    \mathbf{v}_{\rm rem} = \mathbf{v}_{\rm rec, gw} + \mathbf{v}_{\rm rec, ej}.
\end{equation}

We will also consider the magnitudes $v_{\rm rem} = |\mathbf{v_{\rm rem}}|$ of the remnant recoil, and similarly for the recoils coming from the GW emission and ejecta contributions in analysing the results in Sec.~\ref{sec:results}.

\subsection{Calculation of Remnant Mass and Spin}
\label{ssec:mass_spin}
The mass of the remnant is estimated by accounting for the loss of energy due to GW emission [Eq.~(\ref{eq:Erad})], as well as the dynamical mass ejection, i.e.,

\begin{equation}
\label{eq:M_rem}
M_{\rm rem} = M_{\rm ADM} - \int \dot{E}_{\rm GW}~dt -  \int dt \int_{S} \rho~(\mathbf{v} \cdot d\mathbf{S}),
\end{equation}
where $M_{\rm ADM}$ is the initial total binary mass. It is the Arnowitt-Deser-Misner (ADM) mass which is defined as the mass-energy content in a hyperspace [see, e.g., Eq.~(64) of \cite{Gourgoulhon:2000nn}].

While the component neutron stars are non-spinning, the remnant's angular momentum $J_{\rm rem}$ is computed by subtracting the total radiated angular momentum $\Delta J_{GW}$ in gravitational waves from the initial  angular momentum $J_{\rm ADM}$, the ADM angular momentum is defined as the total angular momentum content in a hypersurface [see, e.g., Eq.~(67) of \cite{Gourgoulhon:2000nn}]. For reference also see Eqs.~(2) and (4) in~\cite{Damour:2011fu}. The dimensionless spin of the remnant is then given as 
\begin{equation}
\label{eq:chi_rem}
\chi_{\rm rem} = \frac{cJ_{\rm rem}}{GM_{\rm rem}^{2}}. 
\end{equation}
We note that $J_{\rm rem}$ includes all the remnant, including the disk's angular momentum. As such, $\chi_{\rm rem}$ does not correspond to the spin of the central object alone and it only represents the angular momentum of the remnant that remains after the emission of GWs, while the remnant's spin may be further influenced by dynamical interactions with the disk which are not considered here. Additionally, we do not account for angular momentum loss due to dynamical ejecta since the data available is restricted to the radial velocities of the outflow. Hence, the $\chi_{\rm rem}$ values presented in this paper should be viewed as an upper bound on the remnant's spin.

\section{Results}
\label{sec:results}

Based on the calculations described above, we present the remnant recoil velocity, mass, and spin for 84 unique BNS simulations (as outlined in Sec.~\ref{sec:data}) in our dataset. 
We provide a table in Appendix~\ref{appdx} for the complete list of BNS parameters and corresponding remnant properties.

\subsection{Remnant Recoil Velocity}
\begin{figure}[ht!]
\centering
\subfloat{
\includegraphics[width=0.95\linewidth]{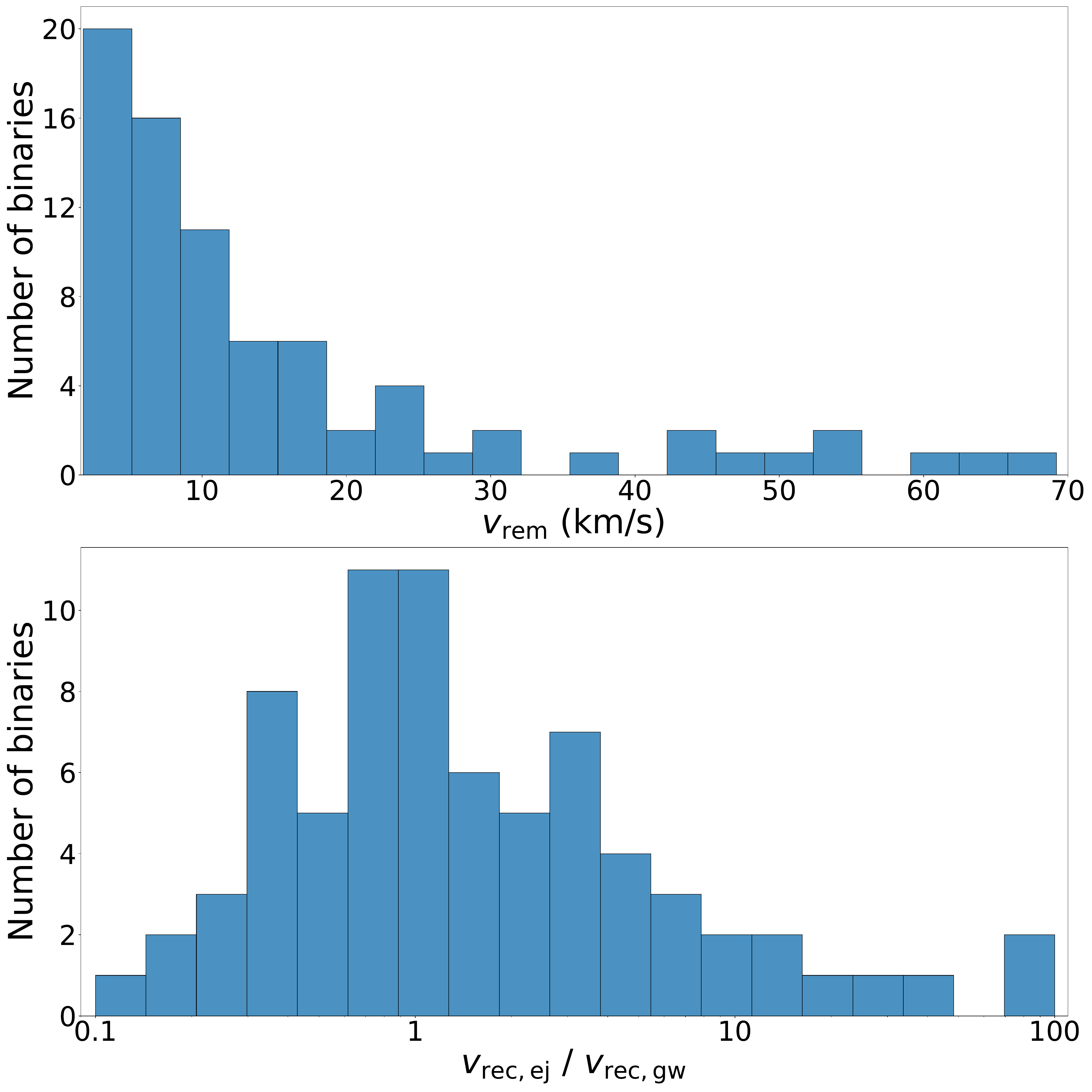}}
\caption{\label{fig:recoil_dist} The distribution of the net remnant recoil velocity (top panel) and the ratio of ejecta to GW recoil velocity (bottom panel). Six binaries with remnant recoils greater than $70$ km/s are not shown in the top panel histogram; refer to Tab.~\ref{table:high_kicks} for their details.
}
\end{figure}

\setlength{\tabcolsep}{9pt} 
\renewcommand{\arraystretch}{2} 
\begin{table*}[t] 
\begin{tabular}{ccccccc} 
\hline 
$q$ & $M (M_{\odot})$ & $\tilde{\Lambda}$ & EoS & $v_{\rm rem}$ 
 (km/s) & $v_{\rm rec, gw}$ 
 (km/s) & $v_{\rm rec, ej}$ 
 (km/s)  \\ 
\hline \hline 
0.89 & 3.31 & 155.93 & BLh & 95.71 & 2.58 & 95.74 \\ 
\hline 
0.55 & 2.88 & 356.96 & SLy4 & 114.07 & 1.54 & 115.33 \\ 
\hline 
0.88 & 2.73 & 360.73 & SLy4 & 125.52 (83.73) & 25.92 (28.72) & 104.51 (69.61) \\ 
\hline 
0.60 & 2.84 & 232.67 & LS220 & 130.38 (135.72) & 1.55 (1.86) & 129.37 (135.10) \\ 
\hline 
0.55 & 2.88 & 200.46 & BLQ & 144.41 (153.64) & 1.30 (1.69) & 145.53 (152.30) \\ 
\hline 
0.55 & 2.88 & 201.53 & BLh & 146.76 (152.37) & 1.27 (1.99) & 147.84 (150.83) \\ 
\hline 
\end{tabular} 
\caption{\label{table:high_kicks} Binary parameters for systems with net recoils ($v_{\rm rem}$) higher than $70$ km/s. The recoils given in the parentheses represent estimates using the next available lower resolution simulation with the same binary parameters (it turned out that only one lower resolution is available for such cases.)}
\end{table*}

The top panel of Fig.~\ref{fig:recoil_dist} shows the distribution of the net recoil velocities imparted to the BNS merger remnants in our dataset. There are systems with net recoils exceeding $70$ km/s and are not shown in Fig.~\ref{fig:recoil_dist}. We have listed them separately in Tab.~\ref{table:high_kicks}. These systems with large net recoils (as shown in Tab.~\ref{table:high_kicks}) predominantly have smaller mass ratios (more unequal masses) with higher total masses. Four of these binaries also have a small ($\sim 1 M_{\odot}$) mass for the secondary neutron star. 
They also have lower values for the reduced tidal deformability parameter $\Tilde{\Lambda}$ ($\lesssim 360$). In all of these cases, the component of the remnant recoil due to ejecta emission dominates its GW counterpart. This is due to larger amounts of matter ejected (at least $6\times 10^{-3} M_{\odot}$) and larger asymmetries in the ejecta distribution compared to the rest of the dataset. These two effects will be discussed in greater detail further in this section. As observed, the recoils are not always negligible for BNS remnants and can be of the order of hundred km/s. This is contrary to the expectation of low recoils in BNS systems because of their negligible spins and comparable masses. As we shall see later, the dynamical ejecta plays an important role in determining the magnitude of these larger recoils.

In the bottom panel of Fig.~\ref{fig:recoil_dist}, we compare the relative contribution of the GW emission and dynamical ejecta to the net recoil. The ratio of $v_{\rm rec, ej}$ to $v_{\rm rec, gw}$ peaks at $1$ indicating that the two contributions are equivalent for a majority of cases. The distribution is skewed towards larger ratios, with $52\%$ of binaries having larger ejecta recoil than GW recoil, with the ejecta recoil being as much as $100$ times greater in a few cases. The range of GW recoils  varies between $[1.3, 45.7]$ km/s
while ejecta recoils span a wider range of $[0.01, 147.8]$ km/s.

\begin{figure*}[!ht]
\centering
\subfloat{
\includegraphics[width=\linewidth]{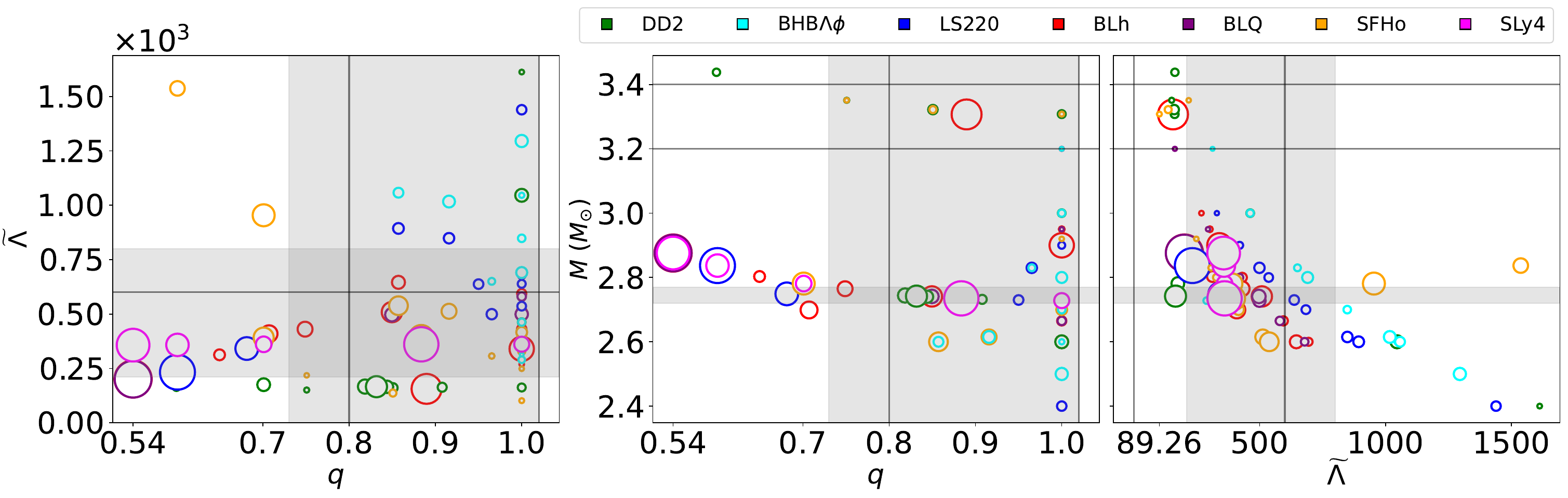}}
\caption{\label{fig:recoils_param_dist} The variation of BNS remnant recoils with binary's mass ratio, total mass, and the reduced tidal deformability parameter. The size of the markers corresponds to the magnitude of the net remnant recoil. The shaded grey region and the region contained within solid grey lines represent the 90\% credible intervals of the respective parameters estimated for GW170817~\cite{LIGOScientific:2017vwq} and GW190425~\cite{LIGOScientific:2020aai}, respectively. For GW170817, we consider parameter estimates using the low-spin prior as described in~\cite{LIGOScientific:2018hze}. Different neutron star EoSs are represented using different colors as indicated in the legend. 
}
\end{figure*}

We now explore the variation of remnant recoils with respect to the intrinsic binary parameters viz. the mass ratio ($q$), total mass ($M$), and the reduced tidal deformability parameter ($\Tilde{\Lambda}$). In Fig.~\ref{fig:recoils_param_dist} we show net recoils as pair-wise 2D scatter plots against these  parameters. The size of the markers corresponds to the magnitude of the net recoil, with larger marker size denoting larger recoils. Different colors represent different EoSs. We observe no obvious correlation of remnant recoils with these binary parameters. However, we note the following patterns: A majority of equal-mass binaries ($q=1$) have low recoils; systems with $\Tilde{\Lambda}$ greater than $1000$ have recoils smaller than $50$ km/s. The larger values of recoils are attained for systems with low mass ratios (more unequal masses) 
and with lower values of $\Tilde{\Lambda}$.

We now take a closer look at aspects that affect the magnitude of the ejecta component of the recoils. There are two main factors at play: (i) the net amount of matter ejected at the end of the merger, and (ii) asymmetries in the distribution of ejecta over a spherical shell around the remnant. Both of these factors play a role in determining how much linear momentum is carried away by the ejecta according to Eq.~(\ref{eq:P_ej}). There is a strong dependence of ejecta recoil on the amount of ejected matter. In our simulations, between $5\times10^{-6}M_{\odot}$ and $0.02 M_{\odot}$ of matter gets ejected from the BNS mergers which accounts for as much as $0.06\%$ of the total mass of these systems. This ejecta is also thrown out at high velocities, as mentioned in Sec.~\ref{ssec:ejecta_recoil}~\cite{Dean:2021gpd, Radice:2018pdn}. 

\begin{figure}[!htb]
\centering
\includegraphics[width=0.95\linewidth]{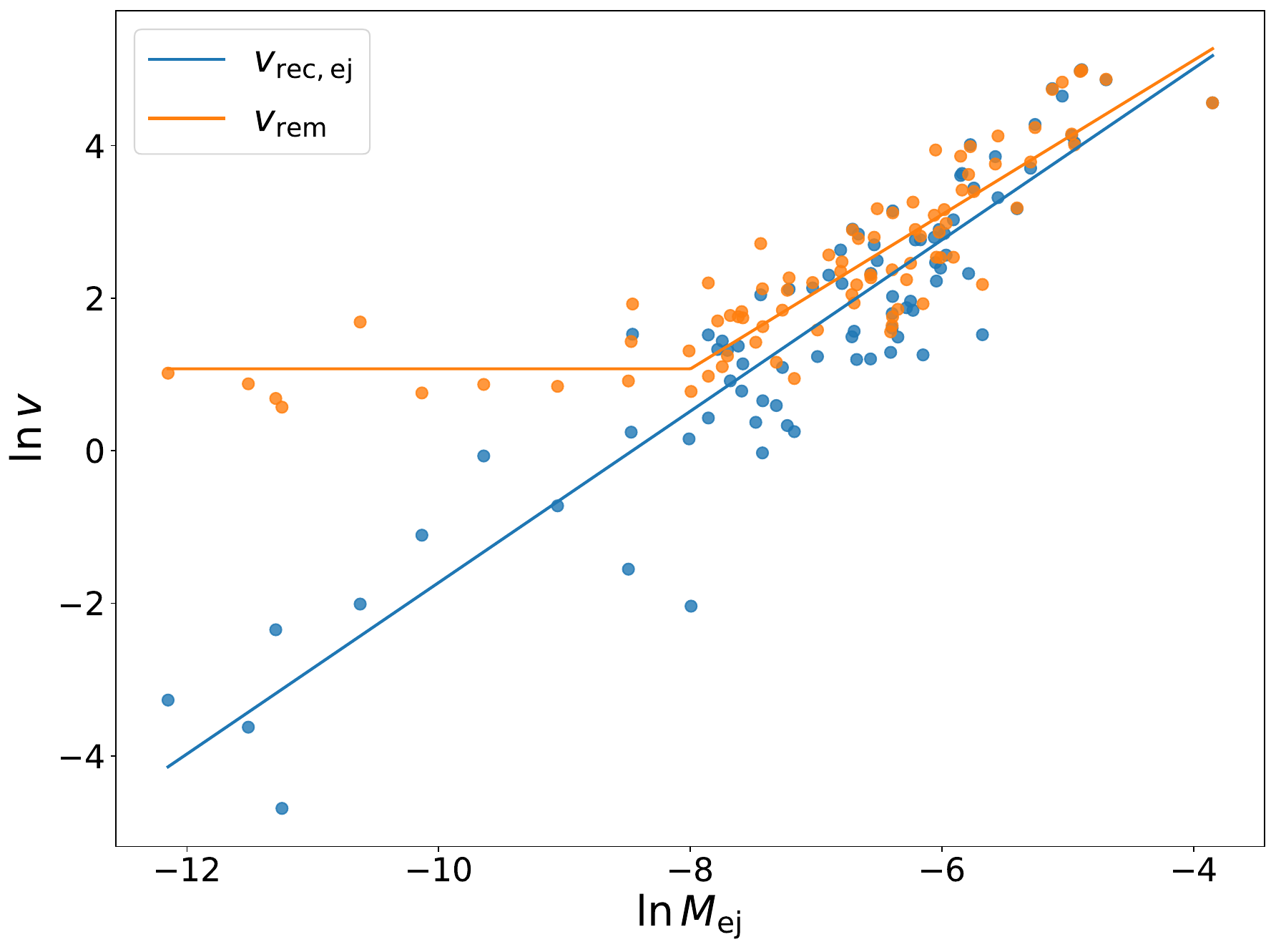}
\caption{\label{fig:recoil_Mej_fit} A scatterplot showing the variation of the ejecta component of the remnant recoil (blue) and the net recoil (orange) with the net ejected mass. Both quantities are correlated and can be fitted with a linear curve (piece-wise linear in the net recoil case) on this logarithmic scale: the solid blue (orange) lines represent the fitting function given in Eq.(~\ref{eq:v_ej_fit}) [Eq.~(\ref{eq:v_rem_fit})]. 
}
\end{figure}

The strong variation of the ejecta recoil with the ejected mass also translates into the net remnant recoil being higher for systems that eject large amounts of matter. In Fig.~\ref{fig:recoil_Mej_fit}, we present a linear fit for ejecta component of the recoil as well as a piece-wise linear fit for the net recoil as a function of the ejected mass. The mathematical expressions for recoils that we obtained after fitting the data are given as
\begin{equation}
\label{eq:v_ej_fit}
    \ln v_{\rm rec, ej} = 1.12^{+0.05}_{-0.05} \ln{M_{\rm ej}} + 9.51^{+0.38}_{-0.38},
\end{equation}
and
\begin{equation}
\label{eq:v_rem_fit}
    \ln v_{\rm rem} = 
    \begin{cases}
        1.075^{+0.15}_{-0.15} & \text{if } \ln{M_{\rm ej}} < -8\\ \\
        1.01^{+0.07}_{-0.07} \ln{M_{\rm ej}} + 9.16^{+0.45}_{-0.45} & \text{if } \ln{M_{\rm ej}} \geq -8,
    \end{cases}
\end{equation}
where $M_{\rm ej}$ is the net amount of the ejected mass in $M_{\odot}$.
We observe that the two curves converge for higher values of $M_{\rm ej}$, implying that the ejecta component of the recoil dominates for mergers that lose a lot of matter. Conversely, ejecta recoils are lower for systems with smaller outflows. For this reason, recoils of binaries at the lower end of $M_{\rm ej}$ are dominated by the GW component, flattening the net recoil distribution that is best described by a constant function. Ideally, this curve should flatten to a zero net recoil, however, we find that it does so at $v_{\rm rem}=2.93$ km/s, which is consistent with the numerical error we get in GW recoils as discussed in Sec.~\ref{sec:discussion}.

\begin{figure*}[ht!]
    \subfloat[Isotropic emission]{%
        \includegraphics[width=.5\linewidth]{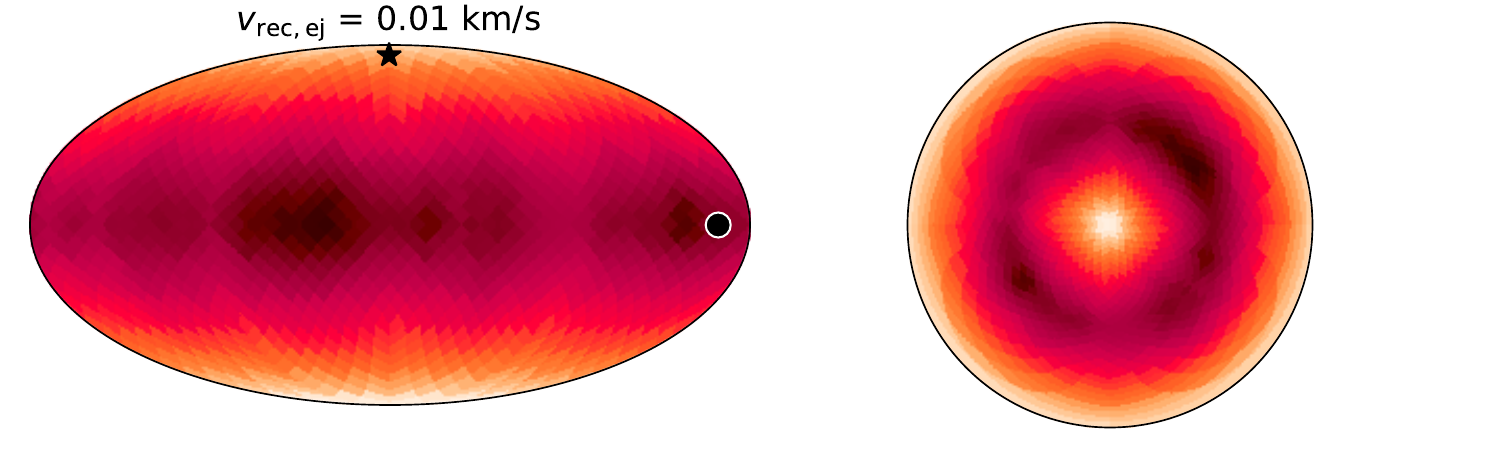}%
        \label{subfig:low_kick_spherical}%
    }
    \subfloat[Symmetric emission]{%
        \includegraphics[width=.5\linewidth]{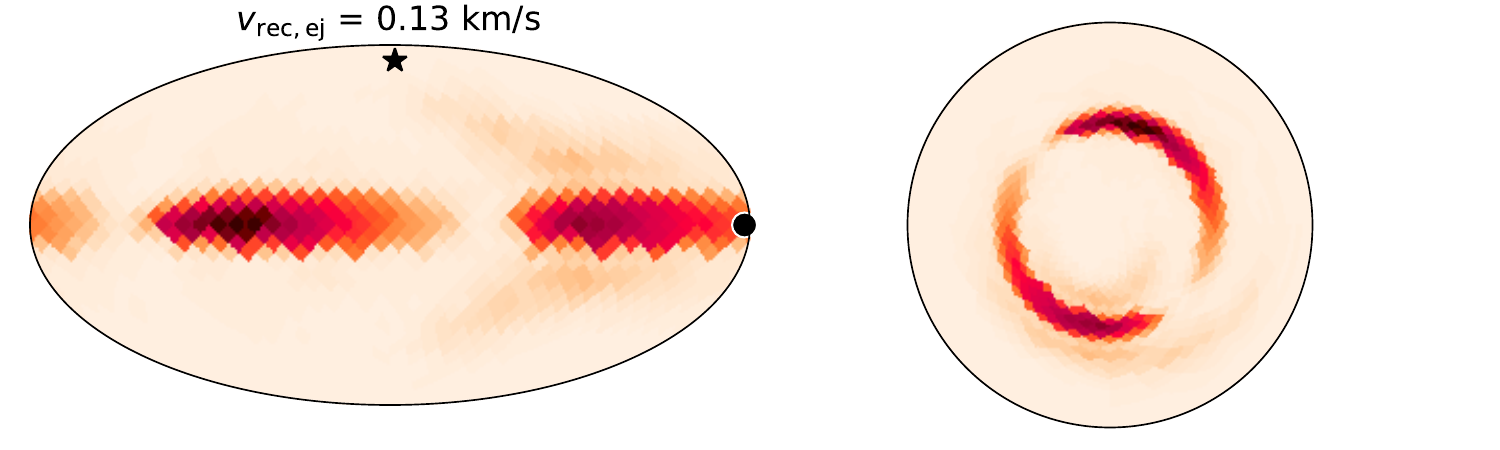}%
        \label{subfig:low_kick_opposite}%
    }\\
    \subfloat[Targeted emission along a fixed direction]{%
        \includegraphics[width=.5\linewidth]{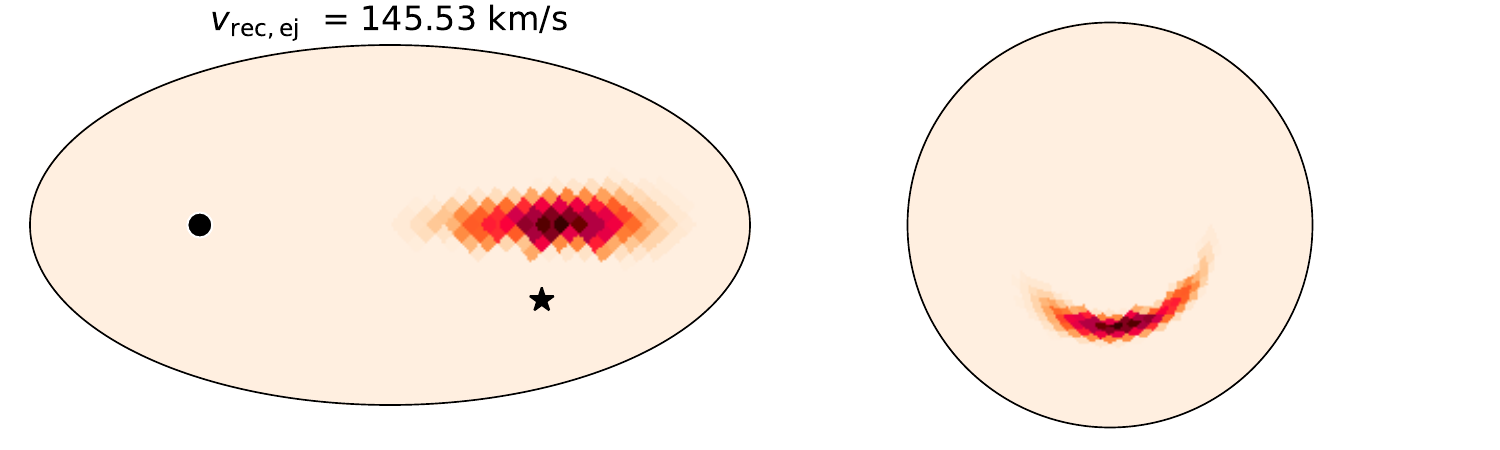}%
        \label{subfig:high_kick_directed}%
    }
    \subfloat[Asymmetric emission with latitudinal spread]{%
        \includegraphics[width=.5\linewidth]{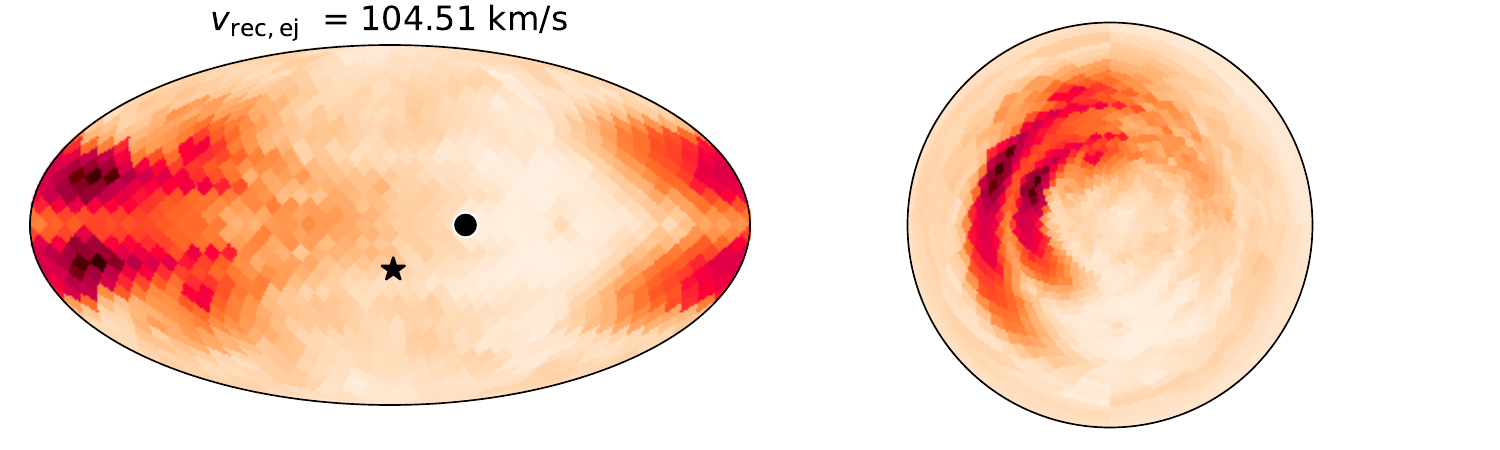}%
        \label{subfig:high_kick_spread}%
    }
    \caption{The distribution of dynamical ejecta across a sphere at $r = 200 M_\odot \sim 295 $ km. 
    around the remnant for cases where the recoils due to ejecta are low (top row) and high (bottom row). The left (right) part of the subplot displays the molleweide (polar) projection of the \textsc{healpix} map. The color intensity represents the ejecta momentum flux density corresponding to each pixel. The star and dot markers indicate the direction of the GW and ejecta components of the remnant recoils respectively.
    }
    \label{fig:ejecta_dist}
\end{figure*}

The second factor that determines the magnitude of the ejecta recoil is how asymmetrically the matter gets ejected. Spherical distributions of the net momentum flux of the ejecta 
can be visualized using maps produced by \textsc{HEALPix}\footnote{http://healpix.sourceforge.net}
and its python library, \textsc{healpy}~\cite{Zonca2019, 2005ApJ...622..759G} as shown in Fig.~\ref{fig:ejecta_dist}. \textsc{HEALPix} provides equal-area pixelization of the 2D spherical shell around the remnant. Here, we map the ejecta distribution into $1200$ equal-area pixels, with each pixel representing the momentum flux  obtained from the mass and velocity distributions at that point. All the velocities, and hence the flux is directed radially outward. Consequently, the localization of flux and its asymmetry over the spherical surface determines the net imbalance of linear momentum when integrated over the whole surface. 

The ejecta distribution over a sphere falls under four categories such as the ones illustrated in subplots of Fig.~\ref{fig:ejecta_dist}. The upper panel has systems exhibiting completely isotropic~(\ref{subfig:low_kick_spherical}): \{$M=3.2 M_{\odot}, q=1.0, \Tilde{\Lambda}=313.0$, EoS=`BHB$\Lambda\phi$'\} or antipodally symmetric~(\ref{subfig:low_kick_opposite}): \{$M=2.6M_{\odot}, q=1.0, \Tilde{\Lambda}=678.8$, EoS=`BLQ'\} ejecta distributions. Here, the momentum flux components cancel out in opposite directions, leading to low ejecta recoils imparted to the remnant. The bottom panel of Fig.~\ref{fig:ejecta_dist} has BNSs with high ejecta recoils due to the momentum flux either being localized to a small region within the sphere (\ref{subfig:high_kick_directed}): \{$M= 2.87 M_{\odot}, q=0.55, \Tilde{\Lambda}=201.53$, EoS=`BLh'\}, or being more spread out and symmetric (\ref{subfig:high_kick_spread}): \{$M= 2.7 M_{\odot}, q=1.0, \Tilde{\Lambda}=416.2$, EoS=`SFHo'\}, but having its magnitude more concentrated in one direction than the other. In either of these two cases, the remnant recoil is high because more ejecta flux components are concentrated in the opposite direction. 

We quantify this asymmetry in ejecta distribution by decomposing it in terms of spherical harmonics. From Eq.~(\ref{eq:P_ej}), we can represent the spherical distribution of the time-averaged ejecta momentum flux density as $F(\theta, \phi)$:

\begin{equation}
\label{eqn:flux_density}
F(\theta, \phi) = \int dt~\rho~v_{r}^{2},
\end{equation}
which can then be expanded in terms of spherical harmonics as:
\begin{equation}
    \label{eqn:flux_decomposition}
    F(\theta,\phi) = \sum_{l=0}^{\infty} \sum_{m=-l}^{l} {a_{lm}Y_{lm}(\theta,\phi)},
\end{equation}
where $Y_{lm}(\theta, \phi)$ are the spherical harmonics.
$F(\theta, \phi)$ is a scalar function since we consider only the radial velocities while computing the dynamical outflow. The complex $a_{lm}$ expansion coefficients can be determined from Eq.~(\ref{eqn:flux_decomposition}) as
\begin{equation}
    a_{lm} = \int_{0}^{2\pi} \int_{0}^{\pi} Y_{lm}^{*}(\theta, \phi) F(\theta,\phi)~\sin{\theta}~d\theta~d\phi.
\end{equation}

We obtain the harmonic decomposition of ejecta momentum flux from the \textsc{healpy} distribution maps using its \textit{map2alm} tool. Using these coefficients $a_{lm}$, we can characterize the asymmetry in the ejecta momentum flux distribution by measuring its antipodal asymmetry~\cite{JAMMALAMADAKA2019436}.

Antipodal symmetry on a sphere is defined by $F(\textbf{x}) = F(-\textbf{x})$ for all $\textbf{x}$ on the sphere.  Under antipodal inversion, $-\textbf{x}(\theta, \phi) \Longrightarrow \textbf{x} (\pi-\theta, \pi+\phi)$, and we have~\cite{JAMMALAMADAKA2019436}:
\begin{equation}
    Y_{l}^{m}(-\textbf{x}) = (-1)^{l} Y_{l}^{m} (\textbf{x}).
\end{equation}
From Eq.~(\ref{eqn:flux_decomposition}), $F(\textbf{x}) = F(-\textbf{x})$ then implies:
\begin{equation*}
    \sum_{l=0}^{\infty} \sum_{m=-l}^{l} a_{lm}Y_{lm}(\pi-\theta, \pi+\phi) = (-1)^{l} \sum_{l=0}^{\infty} \sum_{m=-l}^{l} a_{lm}Y_{lm}(\theta, \phi),
\end{equation*}
which can only be true for even-$l$, i.e., $a_{2l+1}^{m} = 0$ for all $l$ if $F(\textbf{x})$ displays antipodal symmetry.

\begin{figure}[h]
    \includegraphics[width=.95\linewidth]{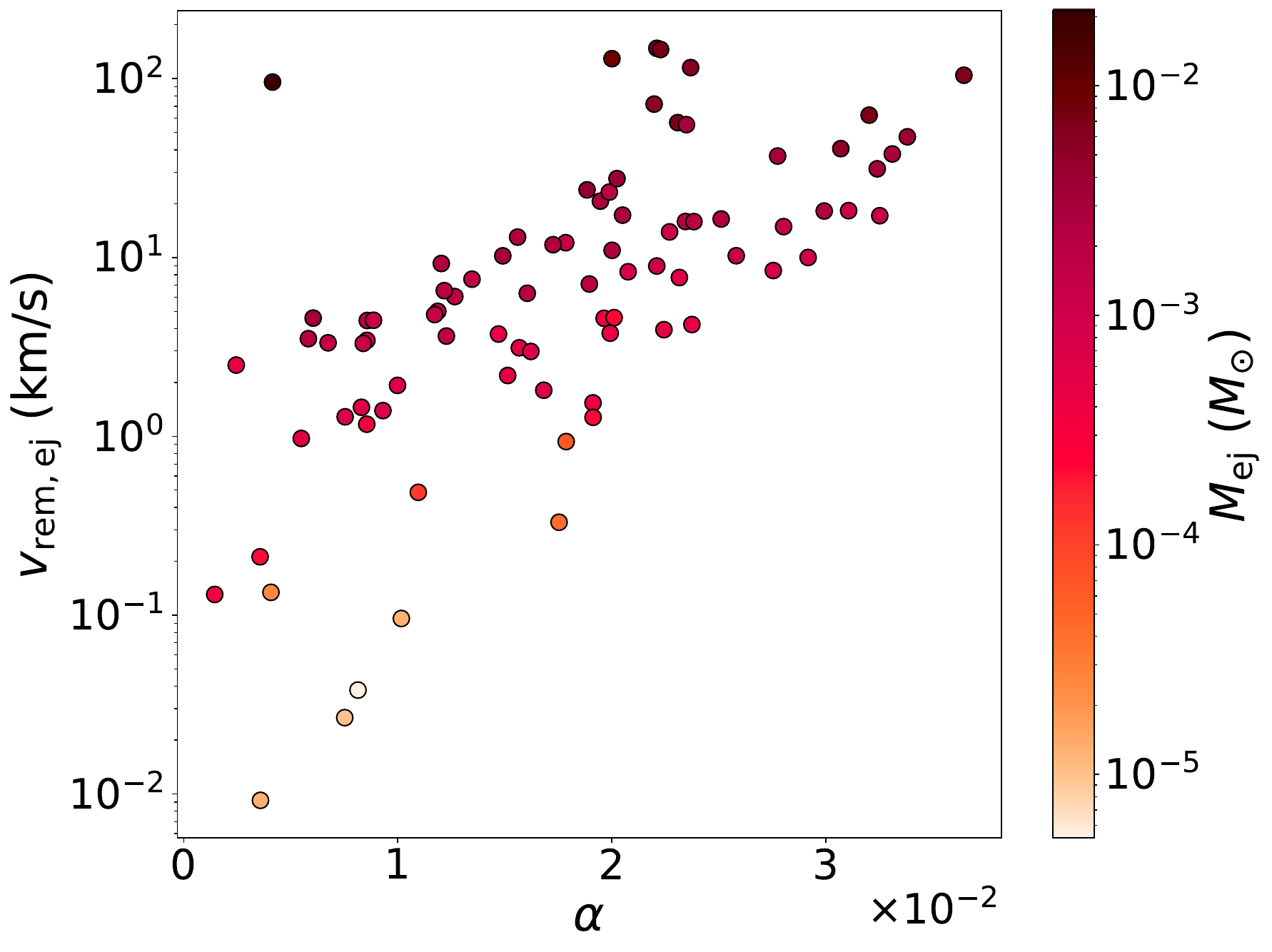}%
    \caption{\label{fig:asymmetries} Quantifying the asymmetry in the ejecta distribution using the antipodal symmetry parameter, $\alpha$. 
    }
\end{figure}

We can define an asymmetry parameter, $\alpha$, by adding all the odd-$l$ coefficients in quadrature:
\begin{equation}
    \alpha = \sqrt{\sum_{\text{odd}\mbox{-}l} a_{lm}^{*}a_{lm}}.
\end{equation}
$\alpha$ is zero for isotropic and antipodally symmetric distributions, with its value increasing as the non-zero odd-$l$ terms contribute to asymmetric distributions. 
Figure~\ref{fig:asymmetries} displays the variation of remnant recoils due to ejecta emission with the asymmetry parameter, $\alpha$. It shows an increase in recoil magnitude with increasing asymmetry. For binaries with similar amounts of $M_{\rm ej}$ indicated by the color, it can be seen that larger asymmetry leads to larger recoils. 

\subsubsection{Comparison with BBH Recoil}\label{subsubsec:comparison}

\begin{figure}[h]
    \includegraphics[width=.95\linewidth]{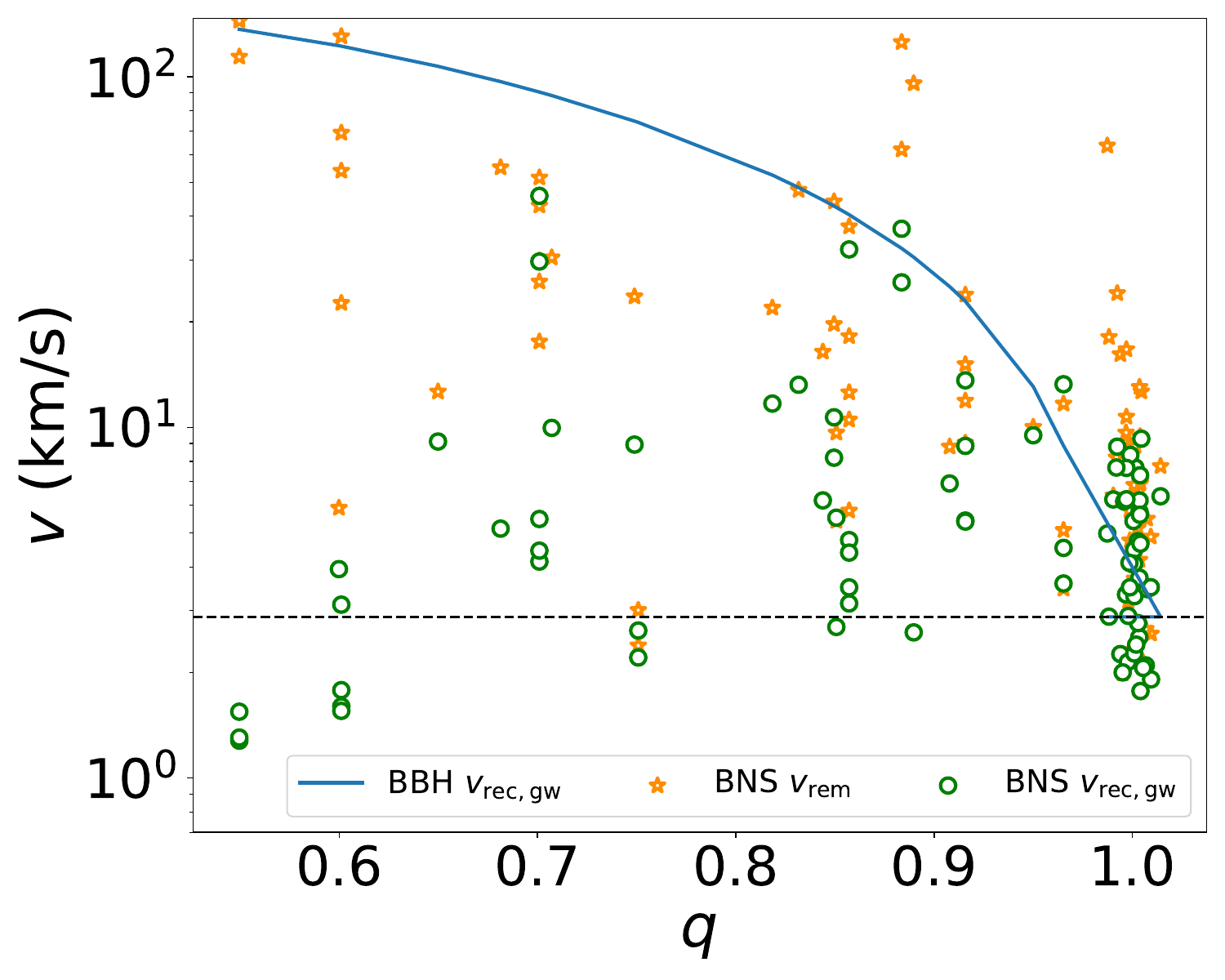}%
    \caption{\label{fig:bbh_comparison} Comparison of BNS remnant recoils with the recoils of non-spinning BBHs having the same mass-ratio. The solid blue line shows the BBH recoils calculated through the GW emission using \textsc{Surrkick}. This line terminates at $v=2.88$ km/s (dashed horizontal line) for $q=1$, which represents numerical errors in the \textsc{surrkick} calculation. The scatter of points represent BNS recoils from our dataset, with green circles
    considering just the BNS's GW recoil while the orange stars considering the net BNS recoil. 
    }
\end{figure}

We compare our NR-based BNS merger remnant recoil with the recoil of a BBH merger remnant of the same mass ratio. To calculate the BBH remnant recoil we use the NRSurrogate fitting code \textsc{surrkick}~\cite{Gerosa2018:PhysRevD.97.104049} where the recoils are obtained solely by considering the binary's GW emission. Figure~\ref{fig:bbh_comparison} shows a comparison of these BBH and BNS recoil velocities. For the BNS kicks, we consider (i) the recoil from GW emission alone (green circles), and (ii) the net recoil including the ejecta contribution (orange stars). 
For equal-mass binaries ($q=1$), we find that the GW component of BNS and BBH recoils are broadly comparable. Note that a non-spinning, equal-mass BBH remnant is expected to have zero kick; but the output of \textsc{surrkick} is $2.88$ km/s in this limit due to numerical errors. 
Whereas in the case of BNS systems, the GW recoils for equalmass and irrotational binaries range from $1.77$ km/s to $9.28$ km/s. These non-zero values unlike in BBH could be attributed to the $m=1$ one-armed spiral instability typically seen in BNS postmerger ~\cite{Radice:2016gym}, in addition to numerical errors reported in Sec.~\ref{sec:discussion}. However, the net recoil for these systems can be even larger due to the ejecta component. As the mass ratio decreases, BBH recoils exceed the GW components of their corresponding BNS counterparts and can be up to two orders of magnitudes greater. But again, the addition of the ejecta component brings the net BNS recoils closer to the BBH values. 

\subsubsection{Equation of State Dependence of BNS recoils}

\begin{figure}[h]
\centering
\subfloat{
\includegraphics[width=0.95\linewidth]{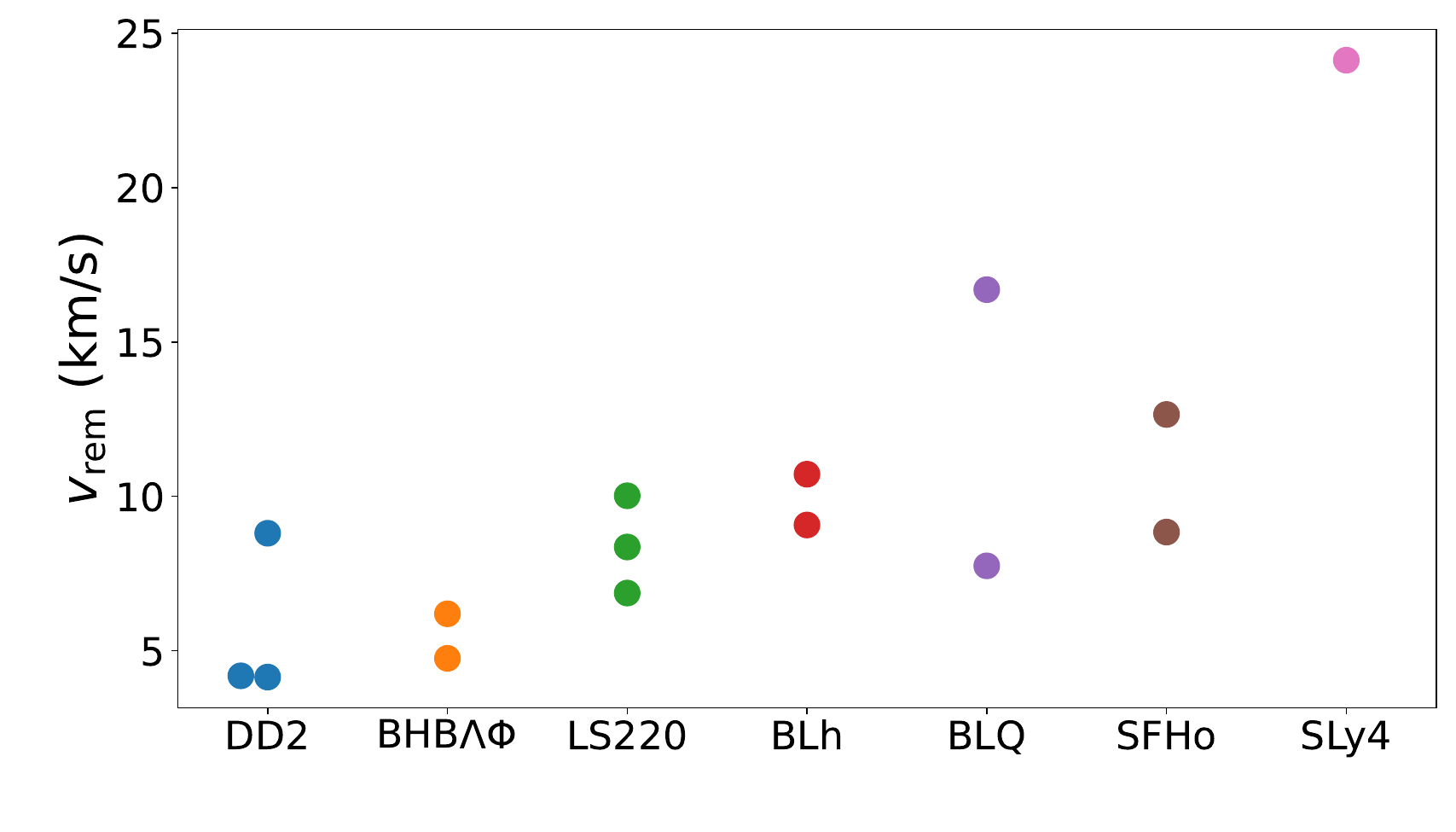}
}
\caption{\label{fig:EoS_recoils} The variation of BNS remnant recoils with the EoS for comparable mass binaries ($q>0.9$) lying in the total mass $M=[2.63,2.75]M_{\odot}$ bin.
}
\end{figure}

As mentioned in Sec.~\ref{sec:data} our NR dataset includes six EoS models (see Fig.~\ref{fig:NS_EOS}).  
In Fig.~\ref{fig:EoS_recoils}, we study the effect of the EoS on the BNS recoil. Unfortunately, in our dataset, we do not have multiple EoS simulations while keeping the binary's other intrinsic parameters (mass ratio, total mass) the same. Hence, to ensure a comparison on equal footing with respect to other binary parameters, we selected simulations lying within the mass ratio bin $[0.9,1.0]$ and total mass bin $M=[2.63,2.75]M_{\odot}$. This bin contains at least one simulation belonging to each EoS. The EoS are ordered from stiffest (\textsc{DD2}) on the left to softest (\textsc{SLy4}) on the right. A softer EoS leads to more dynamical mass ejection which is correlated with larger recoils. We see a similar trend for binaries with ($q<0.9, M<2.63 M_{\odot}$), although there are no simulations belonging to the \textsc{BLQ} and \textsc{SLy4} EoSs in this regime. For binaries in other bins of mass-ratio and total mass, including ($q<0.9$, $M>2.75 M_{\odot}$) bin which has the highest net recoils, we do not find any dependence of BNS kicks on the EoS.

\subsection{Remnant mass and spin}
The remnant masses for the binaries in our dataset varies in the range $[2.34,3.38] M_{\odot}$. The remnant's dimensionless spin ranges from $0.63$ to $0.87$. The remnant spin is evaluated solely based on the angular momentum from the gravitational waveform alone. These values are consistent with the expected spin magnitude of irrotational BNS remnants that are either short-lived or formed by prompt collapse being in the range of $[0.6,0.88]$ as discussed in~\cite{Bernuzzi:2020tgt}. 

\section{Conclusions and Future Directions}
\label{sec:discussion}
In this paper, we presented, for the first time, the numerical relativity estimates of recoil velocities of binary neutron star merger remnants using \textsc{CoRe} database~\cite{Gonzalez:2022mgo}. We computed the recoil due to the emission of gravitational waves as well as dynamical matter ejecta for 84 unique binaries in our dataset. We find that matter ejected asymmetrically in the post-merger kilonova is the leading contributor to the recoil magnitude. Binary neutron star merger recoils can be of the order of $100$ km/s. The recoil magnitudes are generally higher for more unequal mass binaries, reduced tidal deformability parameters below $500$, and systems with the SLy4 equation of state. However, the present dataset proves insufficient in demonstrating clear variations in order to obtain a parameterized fit for recoil velocity for neutron star mergers. We present a fit for the remnant recoil and its ejecta component with the amount of ejected matter, which could be an observable in the electromagnetic follow-up of these mergers~\cite{Villar:2017wcc, Perego:2017wtu}. We compare the gravitational wave kick component and total kick of BNS with the BBH kick for systems having the same mass-ratio.
We also present estimates for remnant mass and spins and found them to be consistent with results in the literature \cite{Zappa:2017xba, Coughlin:2018fis}.

The information on recoil velocities presented in this paper is the first step toward studying the astrophysical environments of binary neutron stars mergers and hierarchical mergers involving low-mass gap black holes \cite{Gupta:2019nwj, Zevin:2020gma, Yang:2020xyi}. 
We found that most of the recoils are less than $50$ km/s in our simulations which implies that such remnant will not leave the dynamical environment such as globular cluster and nuclear star cluster~\cite{Gnedin:2002un, Antonini:2016gqe} as well as the  peripheral regions of galaxies~\cite{Stark:2016dxf}, and may take part in hierarchical mergers The escape velocities of galactic globular clusters peak around 30 km/s with the highest value around 100 km/s \citep{Gnedin:2002un} leading to the escape of some of these systems out of their host environment. The present calculation is required to compute the hierarchical binary neutron star merger rates leading to the fraction of BNS merger product that is retained in a galaxy or a stellar cluster. 
This fraction, a necessary component in the Bayesian inference of the observed stellar-mass BH population~\citep{Gerosa:2021mno}, will benefit from the results of the present work. 
On the other hand, there are binaries in our dataset that exhibit recoils higher than $100$ km/s, this implies that such remnant would leave the habitat post-merger if happened to be in a low escape speed environment. In any case, accurate estimates of recoil velocities of binary neutron star remnants would be crucial in constraining the expected merger rates in different stellar environments~\cite{Ye:2019xvf}. Our comparison of BNS and BBH recoils in Sec.~\ref{subsubsec:comparison} also shows that the dual mechanism of gravitational wave and ejecta emission increases BNS recoils to make them comparable in magnitude to BBH recoils over a range of mass ratios $[0.55,1]$.

Binary neutron star remnant recoils are also crucial in determining the fate of hierarchical triple systems. For instance, the highest asymmetric mass event GW190814~\cite{LIGOScientific:2020zkf} had its secondary close to the low-mass gap with a mass of ${\sim}2.6 M_{\odot}$. One explanation for this system is that the secondary was the remnant of a binary neutron star merger which existed in a triple system with the large black hole~\cite{Lu:2020gfh}. However, a high remnant recoil would in this case disrupt such a triple system from forming GW190814. Another important astrophysical implication of remnant recoils is their impact on the fall-back disk of ejected matter that accretes around the remnant. A remnant moving with a high recoil may cause this disk to be cut or stretched, and could also impact its neutrino and jet emissions~\cite{Kyutoku:2021icp}.

Active Galactic Nuclei (AGN) also provide an environment to efficiently make and evolve binary compact mergers~\cite{Tagawa:2019osr,Tagawa:2020qll}. Such analyses would require accurate knowledge of recoil computed in this work.
Binary mergers in the low mass gap are difficult to explain with our current knowledge of single-star evolution which does not produce black holes in the range $3-5M_\odot$. BNS recoils would be required to compute the merger rates by coupling this information with the large-scale dynamical cluster evolution as shown in \citealt{Rodriguez:2020viw}. In addition, explaining any possible correlation observed in the future between the host environment and properties of BNS mergers must take migration into account.

While estimating recoil velocities presented here, we found that the choice of grid-spacing resolution in the NR simulations induce median relative errors of $25\%$ and $45\%$ in the gravitational wave and ejecta kick components, respectively. These have been calculated based on $55$ simulations for which at least two different resolutions are available. However, a number of high-recoil binaries given in Tab.~\ref{table:high_kicks} show lower percentage errors than the median values. 
As we discussed in Sec.~\ref{sec:data}, contributions from secular ejecta are not included while calculating the recoil velocity. But they are believed to be mostly isotropic and hence will not impact the recoil velocity estimates we have presented in this paper.  

Other processes such as asymmetric jet emission~\cite{Wang:1992_asymmjets} have been proposed to play a part, but are expected to be sub-dominant in affecting the final remnant recoil. We also do not distinguish the effects of hybrid models of equations of state with deconfined quarks \cite{Prakash:2021wpz,Logoteta:2020yxf} and hyperons \cite{Banik:2014qja,Radice:2016rys} and various neutrino transport approximations. Systematic and reliable long-term post-merger simulations are needed to better understand and disentangle the effects of various physics discussed above on ejecta and its properties. While these effects could be important, remnant properties and the recoil velocities presented in this paper can be treated as good estimates for the same which was previously unknown in the literature. 

In the future, we aim to develop a fitting formula for remnant recoil velocity using Gaussian Process Regression techniques as was done in the case of binary black hole systems in~\cite{Varma:2018aht,Varma:2019csw}. We will also estimate the recoil of neutron star - black hole systems using publicly available numerical relativity simulations \cite{Dietrich:2018phi,Kiuchi:2019kzt,Boyle:2019kee}.

\acknowledgements

\label{sec:acknowledgements}
We thank Suvodip Mukherjee for helpful discussions towards analyzing asymmetries in the ejecta momentum flux distributions. We also thank Prasanta Char, Juan Calderon Bustillo, and Nathan Johnson-McDaniel for carefully reading the manuscript and providing valuable comments.
SP acknowledges funding from the U.S. Department of Energy, Office of Science, Division of Nuclear Physics under Award Number(s) DE-SC0021177. AG is supported in part by NSF grants AST-2205920 and PHY-2308887. 
DR acknowledges funding from the U.S. Department of Energy, Office of Science, Division of Nuclear Physics under Award Number(s) DE-SC0021177 and from the National Science Foundation under Grants No. PHY-2011725, PHY-2020275, PHY-2116686, and AST-2108467. Some of the results in this paper have been derived using the \textsc{healpy} and \textsc{HEALPix}~\cite{2005ApJ...622..759G, Zonca2019} packages. 
The software packages NumPy~\cite{Harris:2020xlr}, SciPy~\cite{2020SciPy-NMeth}, AstroPy~\cite{Astropy2018}, Matplotlib~\cite{plt4160265}, and Pandas~\cite{pandas_paper} were utilised for data analysis. 

This is LIGO document number P2300243.

\bibliography{bns_kicksNotes}

\widetext
\appendix
\section{Remnant Properties of BNS Mergers}
\label{appdx}
The following table summarizes the BNS parameters ($q$, $M$, $\Tilde{\Lambda}$, EoS) and properties of their remnants ($M_{\rm rem}$, $\chi_{\rm rem}$, $v_{\rm rem}$) for the 84 binaries used in the analysis presented in this paper.  These simulations are unique to their intrinsic parameters and EoS, after selecting for the highest available NR resolution and the M0 neutrino transport scheme~\cite{Radice:2016dwd}. The \textsc{ID} column corresponds to the name CoRe Database simulation in the publicly available repository~\cite{CoRe_DB_git}. We provide magnitudes of both the GW and ejecta components of the recoil in addition to the magnitude of the net recoil velocity, which is the vector sum of the two. 

\setlength{\tabcolsep}{11pt} 
\renewcommand{\arraystretch}{1.9} 
\begin{longtable*}{cccccccccc} 
\hline 
ID &  $q$ & $M$ & $\tilde{\Lambda}$ & EoS & $v_{\rm rem}$  & $v_{\rm rec, gw}$  & $v_{\rm rec, ej}$  & $M_{\rm rem}$ & $\chi_{\rm rem}$ \vspace{-0.1in} \\ 
 & & $(M_{\odot})$  & & & (km/s) & (km/s) & (km/s) &  $(M_{\odot})$ \\ 
\hline \hline 
THC:0001 & 1.00 & 2.50 & 1295.30 & BHB$\Lambda\Phi$ & 16.16 & 2.26 & 17.13 & 2.45 & 0.80 \\ 
\hline 
THC:0002 & 1.00 & 2.60 & 1045.67 & BHB$\Lambda\Phi$ & 3.19 & 2.90 & 1.81 & 2.54 & 0.77 \\ 
\hline 
THC:0003 & 1.00 & 2.70 & 848.04 & BHB$\Lambda\Phi$ & 6.20 & 5.63 & 2.19 & 2.61 & 0.70 \\ 
\hline 
THC:0004 & 0.92 & 2.62 & 1016.98 & BHB$\Lambda\Phi$ & 15.13 & 8.86 & 7.73 & 2.55 & 0.75 \\ 
\hline 
THC:0005 & 0.86 & 2.60 & 1057.60 & BHB$\Lambda\Phi$ & 10.52 & 4.39 & 13.90 & 2.54 & 0.78 \\ 
\hline 
THC:0006 & 1.00 & 2.80 & 690.63 & BHB$\Lambda\Phi$ & 13.03 & 6.19 & 10.00 & 2.70 & 0.70 \\ 
\hline 
THC:0007 & 0.97 & 2.83 & 650.21 & BHB$\Lambda\Phi$ & 5.09 & 4.53 & 1.93 & 2.74 & 0.71 \\ 
\hline 
THC:0008 & 1.00 & 3.00 & 462.55 & BHB$\Lambda\Phi$ & 5.72 & 4.08 & 3.13 & 2.92 & 0.77 \\ 
\hline 
THC:0009 & 1.00 & 3.20 & 313.05 & BHB$\Lambda\Phi$ & 1.77 & 1.77 & 0.01 & 3.13 & 0.79 \\ 
\hline 
THC:0010 & 1.00 & 2.40 & 1612.25 & DD2 & 2.58 & 1.91 & 1.29 & 2.36 & 0.82 \\ 
\hline 
THC:0011 & 1.00 & 2.50 & 1295.30 & DD2 & 2.66 & 2.10 & 1.54 & 2.45 & 0.81 \\ 
\hline 
THC:0012 & 1.00 & 2.60 & 1045.67 & DD2 & 18.10 & 2.89 & 18.28 & 2.55 & 0.79 \\ 
\hline 
THC:0013 & 1.00 & 2.70 & 848.04 & DD2 & 4.19 & 3.73 & 1.28 & 2.63 & 0.75 \\ 
\hline 
THC:0014 & 0.92 & 2.62 & 1016.98 & DD2 & 9.03 & 5.39 & 4.57 & 2.56 & 0.79 \\ 
\hline 
THC:0015 & 0.86 & 2.60 & 1057.60 & DD2 & 5.79 & 3.15 & 7.56 & 2.55 & 0.81 \\ 
\hline 
THC:0016 & 1.00 & 2.80 & 690.63 & DD2 & 4.88 & 3.50 & 3.44 & 2.72 & 0.74 \\ 
\hline 
THC:0017 & 1.00 & 3.00 & 462.55 & DD2 & 6.32 & 5.41 & 2.98 & 2.90 & 0.68 \\ 
\hline 
THC:0018 & 1.00 & 2.40 & 1439.02 & LS220 & 9.66 & 3.34 & 10.23 & 2.34 & 0.76 \\ 
\hline 
THC:0019 & 1.00 & 2.70 & 683.76 & LS220 & 8.36 & 8.35 & 0.97 & 2.59 & 0.64 \\ 
\hline 
THC:0020 & 0.92 & 2.62 & 848.43 & LS220 & 11.91 & 5.42 & 8.96 & 2.53 & 0.71 \\ 
\hline 
THC:0021 & 0.86 & 2.60 & 893.44 & LS220 & 12.58 & 3.50 & 10.97 & 2.53 & 0.76 \\ 
\hline 
THC:0022 & 1.00 & 2.80 & 536.07 & LS220 & 8.20 & 7.68 & 1.39 & 2.69 & 0.67 \\ 
\hline 
THC:0023 & 0.97 & 2.83 & 499.26 & LS220 & 11.68 & 13.28 & 7.11 & 2.73 & 0.70 \\ 
\hline 
THC:0024 & 1.00 & 2.90 & 420.75 & LS220 & 5.17 & 4.74 & 5.00 & 2.82 & 0.77 \\ 
\hline 
THC:0025 & 1.00 & 3.00 & 330.16 & LS220 & 2.18 & 2.14 & 0.13 & 2.94 & 0.79 \\ 
\hline 
THC:0030 & 1.00 & 2.70 & 416.19 & SFHo & 12.66 & 9.28 & 9.26 & 2.59 & 0.67 \\ 
\hline 
THC:0031 & 0.92 & 2.62 & 512.60 & SFHo & 23.89 & 13.62 & 12.10 & 2.50 & 0.64 \\ 
\hline 
THC:0032 & 0.86 & 2.60 & 538.21 & SFHo & 37.41 & 32.18 & 10.21 & 2.51 & 0.68 \\ 
\hline 
THC:0033 & 1.00 & 2.80 & 328.90 & SFHo & 5.49 & 3.46 & 3.78 & 2.73 & 0.78 \\ 
\hline 
THC:0034 & 0.97 & 2.83 & 307.02 & SFHo & 3.47 & 3.59 & 3.73 & 2.77 & 0.79 \\ 
\hline 
THC:0035 & 1.00 & 2.92 & 248.48 & SFHo & 2.41 & 2.40 & 0.03 & 2.86 & 0.78 \\ 
\hline 
THC:0037 & 0.71 & 2.70 & 409.26 & BLh & 30.51 & 9.97 & 37.91 & 2.69 & 0.78 \\ 
\hline 
THC:0038 & 0.65 & 2.80 & 312.44 & BLh & 12.65 & 9.11 & 20.65 & 2.75 & 0.78 \\ 
\hline 
THC:0039 & 0.55 & 2.88 & 201.53 & BLh & 146.76 & 1.28 & 147.84 & 2.83 & 0.79 \\ 
\hline 
THC:0040 & 0.91 & 2.73 & 163.41 & DD2 & 8.81 & 6.92 & 3.31 & 2.66 & 0.75 \\ 
\hline 
THC:0041 & 0.84 & 2.74 & 165.44 & DD2 & 16.43 & 6.19 & 14.88 & 2.67 & 0.76 \\ 
\hline 
THC:0042 & 0.83 & 2.74 & 165.98 & DD2 & 47.55 & 13.23 & 36.92 & 2.68 & 0.79 \\ 
\hline 
THC:0043 & 0.82 & 2.74 & 166.61 & DD2 & 21.93 & 11.69 & 16.41 & 2.68 & 0.79 \\ 
\hline 
THC:0047 & 0.75 & 3.35 & 150.61 & DD2 & 3.01 & 2.63 & 4.22 & 3.28 & 0.77 \\ 
\hline 
THC:0048 & 1.00 & 3.31 & 162.42 & DD2 & 6.85 & 4.48 & 4.61 & 3.23 & 0.78 \\ 
\hline 
THC:0049 & 0.85 & 3.32 & 136.73 & SFHo & 5.41 & 5.53 & 0.13 & 3.24 & 0.74 \\ 
\hline 
THC:0052 & 1.00 & 2.73 & 510.98 & BLh & 10.72 & 7.66 & 6.05 & 2.62 & 0.65 \\ 
\hline 
THC:0053 & 0.70 & 2.78 & 372.41 & BLh & 29.91 & 4.14 & 31.32 & 2.72 & 0.79 \\ 
\hline 
THC:0054 & 0.60 & 2.84 & 256.06 & BLh & 69.20 & 3.12 & 72.07 & 2.79 & 0.80 \\ 
\hline 
THC:0055 & 0.70 & 2.78 & 175.18 & DD2 & 17.56 & 4.45 & 18.19 & 2.73 & 0.80 \\ 
\hline 
THC:0056 & 1.00 & 2.60 & 678.80 & BLQ & 6.39 & 6.23 & 4.44 & 2.51 & 0.69 \\ 
\hline 
THC:0057 & 1.00 & 2.67 & 580.32 & BLQ & 7.75 & 6.36 & 4.46 & 2.57 & 0.70 \\ 
\hline 
THC:0058 & 1.00 & 2.73 & 499.03 & BLQ & 16.69 & 6.24 & 15.89 & 2.64 & 0.74 \\ 
\hline 
THC:0059 & 1.00 & 2.80 & 420.43 & BLQ & 5.80 & 4.11 & 3.95 & 2.73 & 0.77 \\ 
\hline 
THC:0060 & 1.00 & 2.90 & 331.76 & BLQ & 2.33 & 2.27 & 0.49 & 2.84 & 0.80 \\ 
\hline 
THC:0061 & 0.85 & 2.74 & 496.98 & BLQ & 19.70 & 8.20 & 13.00 & 2.66 & 0.75 \\ 
\hline 
THC:0062 & 1.00 & 3.20 & 163.28 & BLQ & 1.99 & 2.00 & 0.10 & 3.13 & 0.77 \\ 
\hline 
THC:0064 & 0.75 & 2.76 & 430.34 & BLh & 23.61 & 8.93 & 17.25 & 2.68 & 0.73 \\ 
\hline 
THC:0066 & 1.00 & 2.73 & 162.42 & DD2 & 4.15 & 4.07 & 1.45 & 2.66 & 0.76 \\ 
\hline 
THC:0067 & 1.00 & 2.95 & 294.80 & BLQ & 2.13 & 2.06 & 0.33 & 2.89 & 0.80 \\ 
\hline 
THC:0069 & 0.55 & 2.88 & 200.46 & BLQ & 144.41 & 1.30 & 145.53 & 2.83 & 0.79 \\ 
\hline 
THC:0070 & 0.89 & 3.31 & 155.93 & BLh & 95.71 & 2.60 & 95.74 & 3.14 & 0.73 \\ 
\hline 
THC:0071 & 0.60 & 3.44 & 163.44 & DD2 & 5.89 & 3.94 & 2.50 & 3.38 & 0.75 \\ 
\hline 
THC:0072 & 0.85 & 3.32 & 160.71 & DD2 & 9.65 & 2.70 & 8.32 & 3.25 & 0.78 \\ 
\hline 
THC:0073 & 0.75 & 3.35 & 218.21 & SFHo & 2.39 & 2.21 & 0.93 & 3.28 & 0.75 \\ 
\hline 
THC:0074 & 1.00 & 3.31 & 101.72 & SFHo & 2.77 & 2.76 & 0.04 & 3.23 & 0.74 \\ 
\hline 
THC:0076 & 1.00 & 2.73 & 289.85 & BHB$\Lambda\Phi$ & 4.76 & 3.50 & 3.63 & 2.65 & 0.76 \\ 
\hline 
THC:0077 & 0.85 & 2.74 & 508.79 & BLh & 44.07 & 10.69 & 40.61 & 2.64 & 0.70 \\ 
\hline 
THC:0079 & 1.00 & 2.73 & 638.72 & LS220 & 6.87 & 5.74 & 3.52 & 2.63 & 0.70 \\ 
\hline 
THC:0082 & 0.95 & 2.73 & 637.42 & LS220 & 10.02 & 9.50 & 3.33 & 2.61 & 0.64 \\ 
\hline 
THC:0083 & 0.68 & 2.75 & 341.40 & LS220 & 55.13 & 5.14 & 56.78 & 2.71 & 0.79 \\ 
\hline 
THC:0084 & 0.60 & 2.84 & 232.67 & LS220 & 130.39 & 1.55 & 129.37 & 2.79 & 0.82 \\ 
\hline 
THC:0086 & 0.60 & 2.84 & 1537.14 & SFHo & 22.62 & 1.60 & 23.22 & 2.79 & 0.80 \\ 
\hline 
THC:0087 & 0.70 & 2.78 & 953.90 & SFHo & 51.57 & 45.75 & 11.79 & 2.71 & 0.74 \\ 
\hline 
THC:0088 & 1.00 & 2.73 & 395.12 & SFHo & 8.85 & 7.68 & 4.58 & 2.64 & 0.72 \\ 
\hline 
THC:0090 & 0.88 & 2.73 & 394.26 & SFHo & 61.60 & 36.86 & 27.61 & 2.64 & 0.72 \\ 
\hline 
THC:0092 & 0.70 & 2.78 & 391.69 & SFHo & 42.92 & 5.48 & 47.30 & 2.72 & 0.80 \\ 
\hline 
THC:0093 & 1.00 & 2.73 & 361.18 & SLy4 & 24.13 & 8.81 & 23.88 & 2.62 & 0.68 \\ 
\hline 
THC:0095 & 0.88 & 2.73 & 360.73 & SLy4 & 125.53 & 25.93 & 104.51 & 2.63 & 0.68 \\ 
\hline 
THC:0096 & 0.70 & 2.78 & 360.76 & SLy4 & 26.04 & 29.72 & 6.31 & 2.70 & 0.71 \\ 
\hline 
THC:0097 & 0.60 & 2.84 & 358.09 & SLy4 & 53.91 & 1.78 & 55.31 & 2.79 & 0.79 \\ 
\hline 
THC:0098 & 0.55 & 2.88 & 356.96 & SLy4 & 114.07 & 1.55 & 115.33 & 2.83 & 0.78 \\ 
\hline 
THC:0099 & 1.00 & 2.60 & 694.43 & BLh & 6.94 & 4.66 & 4.79 & 2.52 & 0.73 \\ 
\hline 
THC:0100 & 1.00 & 2.67 & 593.93 & BLh & 9.07 & 6.15 & 8.45 & 2.57 & 0.69 \\ 
\hline 
THC:0101 & 0.86 & 2.60 & 645.72 & BLh & 18.19 & 4.77 & 15.87 & 2.53 & 0.75 \\ 
\hline 
THC:0102 & 1.00 & 2.80 & 430.72 & BLh & 9.43 & 7.30 & 6.52 & 2.69 & 0.67 \\ 
\hline 
THC:0103 & 1.00 & 2.90 & 340.16 & BLh & 63.60 & 4.98 & 62.52 & 2.81 & 0.74 \\ 
\hline 
THC:0104 & 1.00 & 2.95 & 302.40 & BLh & 3.70 & 3.30 & 1.17 & 2.88 & 0.79 \\ 
\hline 
THC:0106 & 1.00 & 3.00 & 268.83 & BLh & 2.50 & 2.52 & 0.21 & 2.93 & 0.79 \\ 
\hline 
\end{longtable*}


\end{document}